\documentclass[seceq,preprint]{ptptex}

\usepackage{amssymb,eucal}
\usepackage[dvips]{graphicx}

\newcommand{\wt}{\widetilde}
\newcommand{\wh}{\widehat}
\newcommand{\ol}{\overline}
\newcommand{\del}{\partial}
\newcommand{\ra}{\rightarrow}

\newcommand{\lra}{\leftrightarrow}
\newcommand{\nn}{\nonumber}
\newcommand{\half}{\frac{1}{2}}

\def\tr{\mathop{\rm tr}\nolimits}
\def\Tr{\mathop{\rm Tr}\nolimits}

\newcommand{\AD}[1]{$\ol{\mbox{D~\,}}\!\!\!$#1}
\newcommand{\gym}{g_{Y\!M}}
\newcommand{\cu}{\mathcal{U}}
\newcommand{\cC}{\mathcal{C}}

\newcommand{\cL}{{\cal L}}

\newcommand{\bR}{\mathbb{R}}

\newcommand{\bZ}{\mathbb{Z}}

\newcommand{\cO}{{\cal O}}

\newcommand{\psit}{{\widetilde{\psi}}}
\newcommand{\rhot}{{\widetilde{\rho}}}

\newcommand{\kt}{{\widetilde{k}}}

\newcommand{\wht}{{\widehat{w}}}
\newcommand{\Dht}{{\hat{D}}}

\def\mat#1{\matt[#1]}
\def\matt[#1,#2,#3,#4]{\left(%
\begin{array}{cc} #1 & #2 \\ #3 & #4 \end{array} \right)}

\def\v2#1{\vv2[#1]}
\def\vv2[#1,#2]{\left(%
{#1 \atop #2}\right)}

\newcommand{\Ukk}{U_{\rm KK}}
\newcommand{\Mkk}{M_{\rm KK}}

\newcommand{\UoR}{\left(\frac{\Ukk}{R}\right)}

\newcommand{\delsl}{\partial\!\!\!\slash}
\newcommand{\Dsl}{D \!\!\!\!\slash}
\newcommand{\Dhtsl}{\hat{D} \!\!\!\!\slash}

\notypesetlogo                       

\title{
Low Energy Hadron Physics in Holographic QCD%
}

\author{
Tadakatsu \textsc{SAKAI}$^{1,}$\footnote{E-mail:
tsakai@mx.ibaraki.ac.jp} and Shigeki
\textsc{SUGIMOTO}$^{2,}$\footnote{E-mail:
sugimoto@yukawa.kyoto-u.ac.jp}%
}

\inst{
$^1$Department of Mathematical Sciences, Faculty of Sciences,\\
Ibaraki University, Mito 310-8512, Japan\\
$^2$Yukawa Institute for Theoretical Physics, Kyoto University,\\
Kyoto 606-8502, Japan 
}

\abst{
We present a holographic dual of four-dimensional,
large $N_c$ QCD with massless flavors.
This model is constructed by placing $N_f$ probe D8-branes into a D4
background, where supersymmetry is completely broken.
The chiral symmetry breaking in QCD is manifested as a smooth
interpolation of D8-\AD8 pairs in the supergravity background.
The meson spectrum is examined by analyzing
a five-dimensional Yang-Mills theory that originates from
the non-Abelian DBI action of the probe D8-brane. It is found
that our model yields massless pions, which are identified with
Nambu-Goldstone bosons associated with the chiral symmetry breaking.
We obtain the low-energy effective action of the pion field
and show that it contains the usual kinetic term of the 
chiral Lagrangian and the Skyrme term.
A brane configuration that defines a dynamical baryon 
is identified with the Skyrmion.
We also derive the effective action including the lightest vector
meson. Our model is closely related to that in the hidden local symmetry
approach, and we obtain a Kawarabayashi-Suzuki-Riazuddin-Fayyazuddin-type
relation among the couplings.
Furthermore, we investigate the Chern-Simons term on the probe brane
and show that it leads to the Wess-Zumino-Witten term.
The mass of the $\eta'$ meson is also considered, and we formulate
a simple derivation of the $\eta'$ mass term satisfying
the Witten-Veneziano formula from supergravity.
}

\begin{document}

\maketitle

\section{Introduction}

Recently there have been interesting developments regarding
the gauge theory/string theory duality,
with the aim of obtaining more realistic models
from the phenomenological point of view.
Since the discovery of
 AdS/CFT correspondence (for a review, see Ref.~\citen{adscft}),
the first advance made along this line
was the construction of holographic models
of non-conformal field theories without flavor degrees of freedom
(see Ref.~\citen{noncft} for a review).
A key observation with regard to the construction of holographic
models with flavors was given by Karch and Katz \cite{KK}.
They proposed to incorporate the flavor degrees of freedom in the probe
approximation, where flavor branes are introduced as a probe, so that
the back reaction of the flavor branes is negligible.
This approximation is reliable when $N_f\ll N_c$, 
with $N_{c,f}$ being the number of colors and flavors, respectively.
To this time, this approximation has been applied to various
supergravity (SUGRA) models to study aspects
of large $N$ gauge theories with flavors from the holographic point
of view.\cite{kkw}\tocite{stanislov}
In spite of these developments, it seems that we are still far from a
good understanding of the dynamics of QCD.
For instance, in an interesting paper Ref.~\citen{kmmw}, 
Kruczenski, Mateos, Myers and Winters considered probe D6-branes in a
D4 background, which is supposed to be a dual of four-dimensional
Yang-Mills theory \cite{Witten;thermal}.
On the basis of this model, they explored various aspects of low-energy
phenomena in QCD.
An important ingredient which is still missing from their model,
however, is the appearance of the massless pions as Nambu-Goldstone
bosons associated with the spontaneous breaking of the
$U(N_f)_L\times U(N_f)_R$ chiral symmetry in QCD.

\begin{figure}
\begin{center}
\includegraphics[scale=0.7]{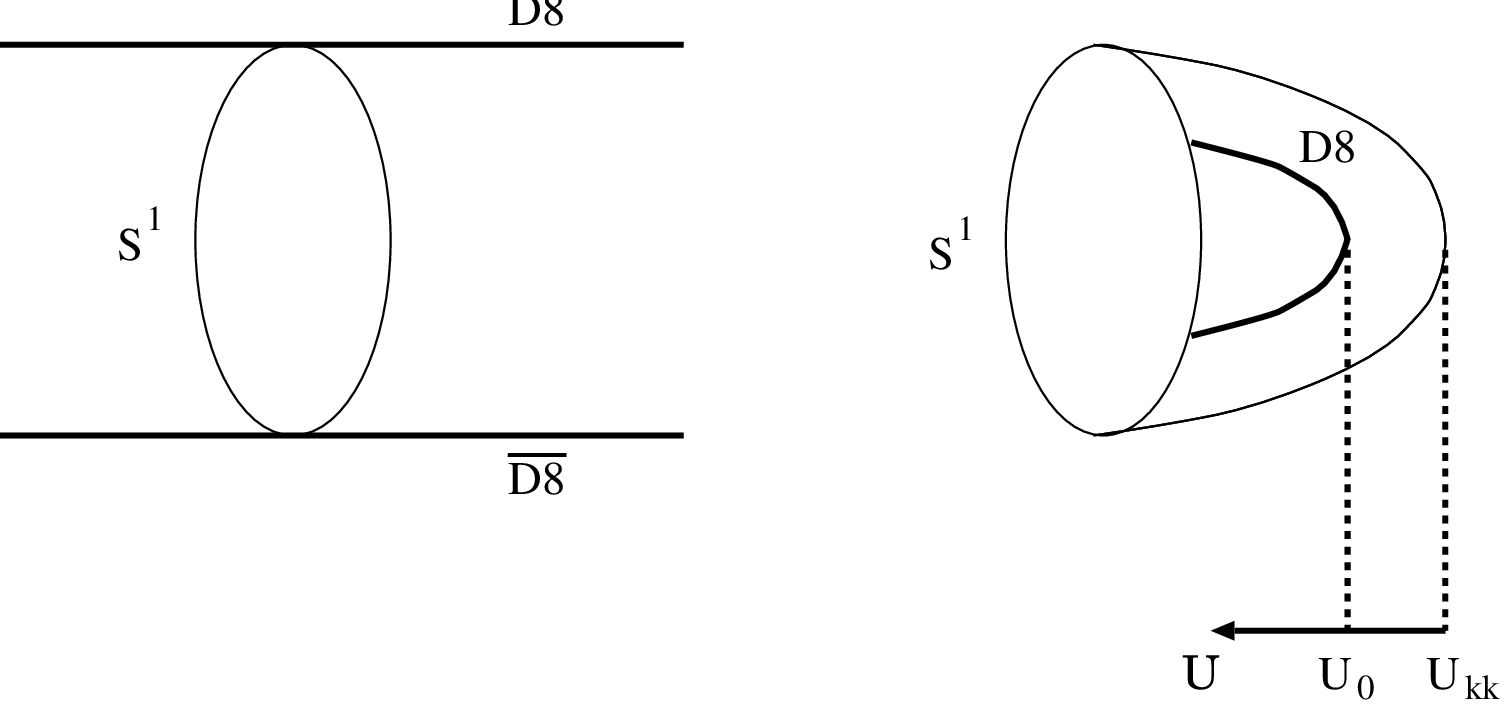}
\end{center}
\caption{\footnotesize{
A sketch of D8 and \AD8 branes.
}}
\label{s1}
\end{figure}

In this paper, we propose another model to
make progress toward a better understanding of QCD with
massless flavors from a holographic point of view.
We construct a holographic model by placing probe D8-branes
into the same D4 background as in Ref.~\citen{kmmw}.
The brane configuration in the weakly coupled regime is given by
$N_c$ D4-branes compactified on a supersymmetry-breaking $S^1$ and
$N_f$ D8-\AD8 pairs transverse to this $S^1$:
\footnote{Holographic descriptions of QCD using
 D4/D8 and D4/D8/\AD8 systems are also considered in Ref.~\citen{Nastase},
 but the brane configuration used there is different from ours.
Specifically, D4 and D8 are parallel in the model used in
 Ref.~\citen{Nastase}, while in our model they are not.}
\begin{eqnarray}
\begin{array}{ccccccccccc}
& 0 & 1 & 2 & 3 & (4) & 5 & 6 & 7 & 8 & 9 \\
\mbox{D4} & \circ & \circ & \circ & \circ & \circ &&&&& \\
\mbox{D8-\AD8}
& \circ & \circ & \circ & \circ &  & \circ & \circ & \circ &\circ & \circ 
\end{array}
\label{D4D8}
\end{eqnarray}
This system is T-dual to the D3/D9/\AD9 system considered in
Ref.~\citen{SugTak}, with one difference being that along the $S^1$
cycle where the T-duality is taken, we impose anti-periodic boundary
conditions for the fermions on the D4-branes in order to break SUSY
and to cause unwanted fields to become massive.
The $U(N_f)_L\times U(N_f)_R$ chiral symmetry in QCD is
realized as the gauge symmetry of the
$N_f$ D8-\AD8 pairs.
The very existence of the compact direction plays a crucial role in
obtaining a holographic understanding of chiral symmetry breaking.
The radial coordinate $U$ transverse to
the D4-branes is known to be bounded from below due to
the existence of a horizon $U\ge \Ukk$ in the supergravity background.
As $U\ra\Ukk$, the radius of the $S^1$ shrinks to zero.
It is found through the study of the DBI action that 
the D8/\AD8 branes merge at some point $U=U_0$
to form a single component of the D8-branes,
yielding, in general, a one-parameter family of solutions
(See Fig. \ref{s1}).\footnote{Similar brane configurations are also
considered in Refs.~\citen{Sakai:2003wu} and \citen{kmmw}.}
On the resultant D8-brane, 
only a single factor of $U(N_f)$ survives as the gauge
symmetry.
We interpret this mechanism as a holographic manifestation of
the spontaneous breaking of the $U(N_f)_L\times U(N_f)_R$
chiral symmetry.

In this paper, we explore various aspects of massless
QCD using the D4/D8 model.
Among those of interest is the appearance of massless pions
as the Nambu-Goldstone bosons associated with the spontaneous chiral
$U(N_f)_L\times U(N_f)_R$ symmetry breaking.
In fact, we find massless pseudo-scalar mesons
in the meson spectrum. We derive the low-energy effective action
of the massless pions and show that it is
identical to that of the Skyrme model,
which consists of the chiral Lagrangian including the Skyrme term
\cite{Skyrme:1}\tocite{anw;skyrme}
(for a review, see Ref.~\citen{skyrme;rev}).
The Skyrme term was introduced by Skyrme to stabilize the soliton
solution in the non-linear sigma model. It was proposed that
this soliton, called a Skyrmion, represent a baryon. We argue that
the Skyrmion can be identified as a D4-brane wrapped around $S^4$,
which is constructed as a soliton in the world-volume gauge theory
of the probe D8-branes. This wrapped D4-brane is nothing but 
Witten's baryon vertex, to which $N_c$ fundamental strings are attached,
and it is considered a color singlet bound state of
the $N_c$ fundamental quarks.\cite{Witten;baryon}

It is not difficult to generalize the effective action to include
vector mesons. We compute three-point couplings involving
the massless pions and the lightest vector mesons.
It is found that the D4/D8 model is closely related to that
employed in the hidden local symmetry approach.
(For a comprehensive review, see Refs.~\citen{bky;review} and
\citen{Meissner}.)
We first discuss how its conceptual framework is interpreted
in the context of the D4/D8 model.
Then, we report the results of a numerical study, from which it is
found that the D4/D8 model exhibits
even quantitatively good agreement with the model used in the
hidden local symmetry approach. In particular,
a Kawarabayashi-Suzuki-Riazuddin-Fayyazuddin (KSRF) type
relation among the couplings is obtained.

We also consider the Chern-Simons (CS) term of the D8-brane.
It is shown to yield the correct chiral anomaly of the
massless QCD. The D4/D8 model enables us to derive the
Wess-Zumino-Witten term in the chiral Lagrangian
through the CS-term of the D8-brane effective action.
We also consider the axial $U(1)_A$ anomaly and its relation to
the mass of the $\eta'$ meson.

Although we expect that our holographic model is in the same universality
class as the four-dimensional massless large $N_c$ QCD,
they are unfortunately not equivalent in the high energy regime,
at least within the supergravity approximation.
Actually, because we obtain a four-dimensional theory by compactifying
D4-branes to a circle of radius $\Mkk^{-1}$, 
an infinite tower of Kaluza-Klein modes of mass scale $\Mkk$ arises.
These Kaluza-Klein modes cannot appear in realistic QCD.
Another difference between the present model and QCD that
we can readily see from the 
brane configuration (\ref{D4D8}) is the existence of the $SO(5)$
symmetry, which acts as the rotation of the $x^5,\cdots,x^9$ plane.
This $SO(5)$ symmetry also appears in the supergravity background
as an isometry. In analyzing the meson spectrum, we mainly focus on
$SO(5)$ singlet states, since such symmetry does not
exist in QCD.\footnote{See Ref.~\citen{PoRuTa} for an argument
for the existence of a decoupling limit in which these exotic
Kaluza-Klein states become much heavier than the states which appear
in four-dimensional QCD.}
A particular class of exotic states
that may be of interest is the fermionic mesons, which
could possibly arise from linear fluctuations of the fermions
on the D8-brane. We investigate the appearance of these states
in Appendix \ref{fermion}.

The organization of this paper is as follows.
In \S \ref{488}, we explain the brane configuration
and the open string spectrum of the D4/D8/\AD8 system,
emphasizing how chiral symmetry emerges.
In \S \ref{probeD8}, we start our analysis with
the $N_f=1$ case. We solve the equation of motion
for the D8-brane DBI action in the D4 background
and pick one solution which is particularly convenient
for the following analysis.
By studying fluctuation modes around the solution, we determine
the degrees of freedom corresponding to the massless pions and
massive vector mesons. 
Numerical analysis of the meson spectrum is given in \S
\ref{numerical}.
In \S \ref{multiNf}, we proceed to the case of multiple
flavors. As long as $N_f\ll N_c$, the probe approximation is still valid.
We derive the Skyrme model as 
the low-energy effective action of the massless pion field
in the D4/D8 model.
The three-point couplings of the pion and the lightest
vector meson are also calculated and
compared with those in the hidden local symmetry
approach. The WZW term is also derived from the CS-term
of the D8-brane in the D4 background.
We also study the baryon configuration and the mass of 
the $\eta'$ meson here.
\S \ref{concdisc} is devoted to a conclusion and discussion.
In Appendix A, we summarize the conventions used for the CS term and
Ramond-Ramond (RR) potentials used in the present paper.
Fluctuations of the fermion on the D8-brane are analyzed
in Appendix \ref{fermion}.

After posting this paper in the preprint archive,
we noticed an important paper, Ref.~\citen{SS}, by Son and Stephanov.
Motivated by the hidden local symmetry approach with a number of
gauge groups, they proposed the idea that a five-dimensional gauge
theory on a curved space is dual to QCD.
Some of the results given in the present paper are given in
Ref.~\citen{SS} as well.
On the other hand, we emphasize that
our model is based on the D4/D8 brane configuration
and the gauge/string correspondence.
We argue that the concept of the hidden local
symmetry is naturally included in our model.
One of the advantages of our model is
that the five-dimensional gauge theory
is fixed by the brane configuration representing $U(N_c)$ QCD
with $N_f$ massless flavors, so that it provides a theoretical
background for the appearance of the five-dimensional
holographic description of QCD.

\section{D4/D8/\AD8 system}
\label{488}
As explained in the introduction, we construct holographic
massless QCD using D4-branes and D8-branes in type IIA string theory.
Before moving to the supergravity description,
let us summarize the open string
spectrum in the weakly coupled regime.
We consider $N_c$ D4-branes and $N_f$ D8-\AD8 pairs
extended as in (\ref{D4D8}). The $x^4$ direction is compactified
on a circle of radius $\Mkk^{-1}$ with anti-periodic boundary
condition for the fermions. For energy scales lower
than $\Mkk$, we effectively obtain a four-dimensional $U(N_c)$ gauge
theory in the D4-brane world-volume.
Due to the boundary condition along the $S^1$, the fermions
that arise from 4-4 strings (the open strings with both ends
attached to the D4-brane) acquire masses of order $\Mkk^{-1}$, and
the supersymmetry is thereby completely broken.
Therefore, the massless modes of the 4-4 strings consist of
the gauge field $A^{(D4)}_\mu$ ($\mu=0,1,2,3$) and the scalar fields
$A^{(D4)}_4$, which is the $x^4$ component of the five-dimensional
gauge field on the D4-brane, and $\Phi^i$ ($i=5,\cdots,9$).
All of these modes belong to the adjoint representation of the gauge
group $U(N_c)$. Because the supersymmetry is broken, the mass terms of
the scalar fields $A^{(D4)}_4$ and $\Phi^i$ are in general produced
via one-loop corrections. The trace part of $A^{(D4)}_4$ and
$\Phi^i$, denoted $a_4$ and $\phi^i$,
are exceptional, and they remain massless, because the mass terms
of these modes are protected by the shift symmetry
$A^{(D4)}_4\ra A^{(D4)}_4+ \alpha 1_{N_c}$
and $\Phi^i\ra \Phi^i+\alpha^i 1_{N_c}$, where
$1_{N_c}$ denotes the $N_c$-dimensional unit matrix.
However, since these modes can only couple to the other
massless modes through irrelevant operators,
we expect that they do not play an important role
in the low-energy physics.

{}From the 4-8 strings and 4-$\ol 8$ strings
(the open strings with one end attached to the D4-brane
and the other end to the D8-brane and \AD8-brane, respectively),
we obtain $N_f$ flavors of massless fermions, which belong to
the fundamental representation of the $U(N_c)$ gauge group.
We interpret these fermions as quarks in QCD.
As discussed in Ref.~\citen{SugTak} for the D3/D9/\AD9 system,
which is T-dual to the present configuration,
the chirality of the fermions created by
4-8 strings is opposite to that created by the 4-$\ol 8$ strings.
Therefore the $U(N_f)_{D8}\times U(N_f)_{\ol{D8}}$ gauge symmetry of
the $N_f$ D8-\AD8 pairs is interpreted as the
$U(N_f)_L\times U(N_f)_R$ chiral symmetry of QCD.

The massless fields on the D4-brane
are listed in Table \ref{massless}, and it is found that we obtain
four-dimensional $U(N_c)$ QCD with $N_f$ flavors with
manifest $U(N_f)_L\times U(N_f)_R$ chiral symmetry.
\begin{table}[htb]
\begin{center}
\parbox{75ex}{
\caption{
\small
The massless fields on the D4-brane with D8-\AD8 pairs.
Here $\textbf{2}_{+}$ and $\textbf{2}_{-}$ denote the
positive and negative chirality spinor representations
of the Lorentz group $SO(3,1)$.
}
\label{massless}
}
$$
\begin{array}{c|cccc}
\hline\hline
\mbox{field} & U(N_c) & SO(3,1) & SO(5) & U(N_f)_L \times U(N_f)_R \\
\hline
A_{\mu}^{(D4)} & \textbf{adj.} & \textbf{4} & \textbf{1} &
 (\textbf{1},\textbf{1})\\
a_{4}  & \textbf{1} & \textbf{1} & \textbf{1} &
 (\textbf{1},\textbf{1})\\
\phi^{i}  & \textbf{1} & \textbf{1} & \textbf{5} &
 (\textbf{1},\textbf{1})\\
\hline
q_{L}^{f}  & \textbf{fund.} & \textbf{2}_+
 &\textbf{1}& (\textbf{fund.},\textbf{1}) \\
q_{R}^{\bar{f}}  & \textbf{fund.}  & \textbf{2}_-
&\textbf{1}& (\textbf{1},\textbf{fund.})\\
\hline
\end{array}
$$
\end{center}
\end{table}

As discussed in Ref.~\citen{SugTak}, we could add a mass term for
the quarks by including the tachyon field created by
the 8-$\ol{8}$ string. However, because we use the DBI action
of the D8-branes to analyze the system in the following sections,
we make the tachyon field massive by separating the D8-branes and
\AD8-branes along the $x^4$ direction, as depicted on
the left-hand side of Fig. \ref{s1}.
More explicitly, the mass of the tachyon mode is given by
\begin{eqnarray}
m^2=\left(\frac{\Delta x^4}{2\pi\alpha'}\right)^2
-\frac{1}{2\alpha'}\ ,
\end{eqnarray}
where $\Delta x^4$ is the distance between the D8-branes
and the \AD8-branes. We can make this mode massive
by choosing $\Delta x^4> \sqrt{2}\pi l_s$.

The massless spectrum on the D4-brane and the minimal couplings among
them are not affected by the separation $\Delta x^4$, as
they do not involve 8-$\ol 8$ strings. Hence,
the low-energy theory is expected to be independent of $\Delta x^4$.

\section{Probe D8-brane}
\label{probeD8}
In order to obtain a holographic
dual of the large $N_c$ gauge theory,
we consider the SUGRA description of the D4/D8/\AD 8 system discussed
in the previous section in the decoupling limit.
Assuming $N_f\ll N_c$, we treat
the D8-\AD8 pairs as probe D8-branes
embedded in the D4 background.
In this section, we present the essential ingredients
employed in subsequent sections by analyzing the case with one flavor.
Generalization to the multi-flavor case is
given in \S \ref{multiNf}.

\subsection{The D4 background}

The D4 background we consider here consists of $N_c$
flat D4-branes with one of the spatial world-volume
directions compactified on $S^1$, along which anti-periodic
boundary conditions are imposed for fermions.
This background yields a holographic dual of four-dimensional
pure Yang-Mill theory at low energies.\cite{Witten;thermal}
Here we mainly employ the notation used in Ref.~\citen{kmmw}
and summarize the relation between the parameters
in the supergravity solution and in the gauge theory.

The D4-brane solution reads
\begin{eqnarray}
&&ds^2=\left(\frac{U}{R}\right)^{3/2}
\left(\eta_{\mu\nu}dx^\mu dx^\nu+f(U)d\tau^2\right)
+\left(\frac{R}{U}\right)^{3/2}
\left(\frac{dU^2}{f(U)}+U^2 d\Omega_4^2\right),
\nn\\
&&~~~~e^\phi= g_s \left(\frac{U}{R}\right)^{3/4},
~~F_4=dC_3=\frac{2\pi N_c}{V_4}\epsilon_4 \ ,
~~~f(U)=1-\frac{\Ukk^3}{U^3} \ ,
\label{D4sol}
\end{eqnarray}
where $x^\mu$ ($\mu=0,1,2,3$) and $\tau$ are the directions
along which the D4-brane is extended.
$d\Omega_4^2$, $\epsilon_4$ and $V_4=8\pi^2/3$
are the line element, the volume form and the volume of a unit $S^4$,
respectively. 
 $R$ and $\Ukk$ are constant parameters. $R$ is related
to the string coupling $g_s$ and string length $l_s$ as
$R^3=\pi g_s N_c l_s^3$.

The coordinate $U$ is bounded from below by the condition $U\ge \Ukk$.
In order to avoid a singularity at $U=\Ukk$,
$\tau$ must be a periodic variable with
\begin{eqnarray}
\tau\sim\tau+\delta\tau \ ,~~~~\delta\tau\equiv\frac{4\pi}{3}
\frac{R^{3/2}}{\Ukk^{1/2}} \ .
\end{eqnarray}
We define the Kaluza-Klein mass as
\begin{equation}
\Mkk=\frac{2\pi}{\delta\tau}=\frac{3}{2}\frac{\Ukk^{1/2}}{R^{3/2}} \ ,
\end{equation}
which specifies the energy scale below which the dual
gauge theory is effectively the same as
four-dimensional Yang-Mills theory.
The Yang-Mills coupling $\gym$ at
the cutoff scale $\Mkk$ can be read off of
the DBI action of the D4-brane compactified on the
$S^1$ as $\gym^{2}=(2\pi)^2 g_s l_s/\delta\tau$.
The parameters $R$, $\Ukk$ and $g_s$ are expressed
in terms of $\Mkk$, $\gym$ and $l_s$ as
\begin{eqnarray}
R^3=\half\frac{\gym^2 N_c l_s^2}{\Mkk} \ ,~~~
\Ukk=\frac{2}{9}\gym^2 N_c \Mkk l_s^2 \ ,~~~
g_s=\frac{1}{2\pi}\frac{\gym^2}{\Mkk l_s} \ .
\label{RUg}
\end{eqnarray}
As discussed in Ref.~\citen{kmmw}, this supergravity description
is valid in the case $1\ll \gym^2 N_c \ll 1/\gym^4$.

\subsection{Probe D8-brane configuration}

Here we consider a D8-brane probe in the D4 background,
which corresponds to the D4/D8/\AD8 system discussed
in \S \ref{488}. The analysis here is parallel to
that given in Ref.~\citen{kmmw} for the D4/D6/\AD6 system.

We consider a D8-brane embedded in the D4 background
(\ref{D4sol}) with $U=U(\tau)$. Then the induced metric
on the D8-brane is given by
\begin{eqnarray}
&&ds^2_{D8}=\left(\frac{U}{R}\right)^{3/2}
\eta_{\mu\nu}dx^\mu dx^\nu+
\left(
\left(\frac{U}{R}\right)^{3/2}f(U)
+\left(\frac{R}{U}\right)^{3/2}\frac{{U'}^2}{f(U)}
\right)d\tau^2
+\left(\frac{R}{U}\right)^{3/2}U^2 d\Omega_4^2 \ ,\nn\\
\end{eqnarray}
with $U'=\frac{d}{d\tau}U$.
The D8-brane action is proportional to
\begin{eqnarray}
S_{D8}
&\propto&
\int d^4x d\tau\, \epsilon_4 \,e^{-\phi} \sqrt{-\det(g_{D8})}\nn\\
&\propto&\int d^4x d\tau\, U^4 \sqrt{f(U)+
\left(\frac{R}{U}\right)^{3} \frac{U'^2}{f(U)}} \ .
\label{D8Utau}
\end{eqnarray}
Since the integrand of (\ref{D8Utau}) does not
explicitly depend on $\tau$, we can obtain 
the equation of motion as the energy conservation law,
\begin{eqnarray}
\frac{d}{d\tau}\left(
\frac{U^4 f(U)}{\sqrt{
f(U)+\left(\frac{R}{U}\right)^{3} \frac{U'^2}{f(U)}}
}\right)=0 \ .
\label{eom;probe}
\end{eqnarray}
The form of the solution of this equation
is depicted in the right-hand side of
Fig. \ref{s1}. Assuming the initial conditions
$U(0)=U_0$ and $U'(0)=0$ at $\tau=0$,
the solution of this equation of motion is obtained as
\begin{eqnarray}
\tau(U)=U_0^4 f(U_0)^{1/2}\int_{U_0}^U
\frac{dU}{\left(\frac{U}{R}\right)^{3/2}
f(U)\sqrt{U^8 f(U)-U_0^8 f(U_0)}} \ .
\end{eqnarray}
The qualitative features of this solution
are similar to those found in Ref.~\citen{kmmw} for
the D4/D6/\AD6 system.
It can be shown that the asymptotic value of
$\tau(U)$ in the limit $U\ra\infty$
is a monotonically decreasing function of $U_0$,
which varies from
 $\tau(\infty)|_{U_0=\Ukk}=\delta\tau/4$
to $\tau(\infty)|_{U_0\ra\infty}=0$ \ .
When $U_0=\Ukk$, the D8-brane and the \AD8-brane
are at antipodal points on the $S^1$ parameterized by
$\tau$. In fact, $\tau(U)=\delta\tau/4$ is a solution of
(\ref{eom;probe}) with the correct boundary conditions.
As $U_0\ra\infty$, a D8-\AD8 pair is sent to infinity and
disappears.

The physical interpretation of this behavior
in the dual gauge theory
description is not clear to us.
The asymptotic value of $\tau$ should correspond to
 $\Delta x^4$ appearing in \S \ref{488}.
However, because, as argued in \S \ref{488}, $\Delta x^4$
is expected to be irrelevant in the low-energy effective
theory on the D4-brane, it is
peculiar that the D8-brane configuration
depends strongly on the asymptotic value of $\tau$.
We leave this issue as a future work.

In what follows, we concentrate on the case $U_0=\Ukk$.
In this case, it is useful to introduce the
new coordinates $(r,\theta)$ or $(y,z)$ in place of
$(U,\tau)$ with the relations
\begin{eqnarray}
y=r \cos\theta \ ,~~~~z=r\sin\theta \ ,
\end{eqnarray}
and
\begin{eqnarray}
U^3=\Ukk^3+\Ukk r^2,~~~
\theta \equiv \frac{2\pi}{\delta\tau}\,\tau=
\frac{3}{2}\frac{\Ukk^{1/2}}{R^{3/2}}\,\tau  \ .
\end{eqnarray}
Then the metric in the $(U,\tau)$ plane is written
\begin{eqnarray}
ds^2_{(U,\tau)} &=&
\left(\frac{U}{R}\right)^{3/2}f(U)\,d\tau^2
+\left(\frac{R}{U}\right)^{3/2}\frac{dU^2}{f(U)}\nn\\
&=&\frac{4}{9}\left(\frac{R}{U}\right)^{3/2}
\left(\frac{\Ukk}{U} dr^2+r^2 d\theta^2\right)\ ,
\label{r-theta}
\end{eqnarray}
with
\begin{eqnarray}
\frac{\Ukk}{U} dr^2+r^2 d\theta^2
=(1-h(r)z^2)dz^2+(1-h(r)y^2) dy^2-2 h(r) zydzdy \ ,
\end{eqnarray}
where $h(r)=\frac{1}{r^2}\left(1-\frac{\Ukk}{U}\right)$.
Note that (\ref{r-theta}) approaches the metric of a flat
two-dimensional plane near $U=\Ukk$ and also that
$h(r)$ is a regular non-vanishing function
in the neighborhood of $r=0$.

Now we consider a D8-brane extending along the $x^\mu$ ($\mu=0,1,2,3$)
and $z$ directions and wrapped around the $S^4$. The position of
the D8-brane in the $y$ direction is denoted  $y=y(x^\mu,z)$.
As we have seen above, $y(x^\mu,z)=0$ is a solution of the
equations of motion of the D8-brane world-volume theory. 
In order to show the stability of the probe configuration, 
let us next examine small fluctuations around this solution.
The induced metric on the D8-brane is now written
\begin{eqnarray}
ds^2
&\!=\!&
\frac{4}{9}\left(\frac{R}{U}\right)^{3/2}\,
\left[\left(\frac{\Ukk}{U}+\dot y^2+h(z)(y^2-2zy\dot y)\right)dz^2
+2\left(\dot y- h(z) zy
\right)\del_\mu y\, dx^\mu dz \right]+\nn\\
&&+\left(\frac{U}{R}\right)^{3/2}
\left(
\eta_{\mu\nu}+\frac{4}{9}\left(\frac{R}{U}\right)^{3}
\del_\mu y\del_\nu y
\right)dx^\mu dx^\nu
+R^{3/2}U^{1/2} d\Omega_4^2+\cO(y^4)\nn\\
&\equiv& ds_{(5\dim)}^2+ R^{3/2}U^{1/2} d\Omega_4^2 \ ,
\end{eqnarray}
where $\dot{y} =\del_zy$
and we have used the identity $1-h(r) z^2=h(r) y^2 +\frac{\Ukk}{U}$.
Then, using the formula
\begin{eqnarray}
\det\mat{A,B,C,D}=\det A \cdot\det(D-CA^{-1}B) \ ,
\end{eqnarray}
and ignoring $\cO(y^4)$ terms, we have
\begin{eqnarray}
&&\sqrt{-\det g_{\rm (5 dim)}}\nn\\
&\simeq&\frac{2}{3}\left(\frac{U}{R}\right)^{9/4}
\left(\frac{\Ukk}{U}\right)^{1/2}
\left(
1+\frac{2}{9}\left(\frac{R}{U}\right)^{3}\eta^{\mu\nu}
\del_\mu y\del_\nu y +\frac{U}{2\Ukk}
\left(h(z)(y^2-2zy\dot y)+\dot y^2\right)
\right) .\nn\\
\end{eqnarray}
Plugging this into the DBI action of the D8-brane, we obtain
\begin{eqnarray}
S_{D8}&=&- T\int d^4x dz \,\epsilon_4\,
e^{-\phi}\sqrt{-\det g_{\rm (5dim)}} \left(R^{3/2}U^{1/2}\right)^2 \nn\\
&\simeq&-\wt T\int d^4x dz \,\left[\,U_z^2
+\frac{2}{9}\frac{R^3}{U_z}\eta^{\mu\nu}\del_\mu y\del_\nu y+
y^2+\frac{U_z^3}{2\Ukk}\dot y^2\,
\right]
\label{fluc;y}
\end{eqnarray}
up to quadratic order in $y$,
where we have defined
$\wt T\equiv\frac{2}{3}R^{3/2}\Ukk^{1/2} \, T V_{4}\,g_s^{-1}$, with
$T=1/((2\pi)^8l_s^9)$, and
\begin{eqnarray}
U_z(z)\equiv (\Ukk^3+\Ukk z^2)^{1/3} \ .
\end{eqnarray}
Here the $U(1)$ gauge potential on the D8-brane
is omitted for simplicity.
Then the energy density carried by a fluctuation
along the $y$ direction is
\begin{eqnarray}
{\cal E}\simeq\wt T \int dz \left[\,
\frac{2}{9}\frac{R^3}{U_z}
\sum_{\mu=0}^3
(\del_\mu y)^2+
y^2+\frac{U_z^3}{2\Ukk}\dot y^2\,
\right]\ge 0 \ .
\end{eqnarray}
This guarantees that the probe configuration we found is stable
with respect to small fluctuations.

\subsection{Gauge field}
\label{fluc}

In this subsection,
we consider the gauge field on the probe D8-brane configuration
considered in the previous subsection.
The gauge field on the D8-brane has nine components,
$A_\mu$ ($\mu=0,1,2,3$), $A_z$ and $A_\alpha$
 ($\alpha=5,6,7,8$, the coordinates on the $S^4$).
As mentioned in the introduction, we are mainly
interested in the $SO(5)$ singlet states, and here
we set $A_\alpha=0$ and assume that $A_{\mu}$ and $A_z$ 
are independent of the coordinates on the $S^4$.
Then the action becomes
\begin{eqnarray}
S_{D8}&=&
-T\int d^9 x \,e^{-\phi}
\sqrt{-\det(g_{MN}+2\pi\alpha' F_{MN})}+S_{CS}\nn\\
&=&-\wt T(2\pi\alpha')^2
\int d^4x dz\, \left[\,
\frac{R^3}{4U_z}
\eta^{\mu\nu}\eta^{\rho\sigma} F_{\mu \rho}F_{\nu \sigma}
+\frac{9}{8}\frac{U_z^3}{\Ukk}
\eta^{\mu\nu} F_{\mu z}F_{\nu z}
\right]+\cO(F^3) \ .
\label{FF}
\end{eqnarray}

Let us assume that $A_\mu$ ($\mu=0,1,2,3$) and $A_z$ can
be expanded in terms of
complete sets $\{\psi_n(z)\}$ and $\{\phi_n(z)\}$,
whose ortho-normal conditions will be specified below, as
\begin{eqnarray}
A_\mu(x^\mu,z)&=&\sum_{n} B_\mu^{(n)}(x^\mu) \psi_n(z) \ ,
\label{expand;Av}\\
A_z(x^\mu,z)&=&\sum_{n} \varphi^{(n)}(x^\mu) \phi_n(z) \ .
\label{expand;A}
\end{eqnarray}
The field strengths are given by
\begin{eqnarray}
F_{\mu\nu}(x^\mu,z)&=&\sum_n \left(\del_\mu B^{(n)}_\nu(x^\mu)
-\del_\nu B^{(n)}_\mu(x^\mu)\right)\psi_n(z)\nn\\
&\equiv&\sum_n F_{\mu\nu}^{(n)}(x^\mu)\psi_n(z) \ ,\\
F_{\mu z}(x^\mu,z)&=&\sum_n \left(\del_\mu
\varphi^{(n)}(x^\mu)\phi_n(z) -B^{(n)}_\mu(x^\mu)\dot\psi_n(z)\right).
\label{F}
\end{eqnarray}
Then, the action (\ref{FF}) is written
\begin{eqnarray}
S_{D8}
&=&-\wt T(2\pi\alpha')^2
\int d^4x dz\, \sum_{m,n}
\left[\,
\frac{R^3}{4U_z}
F_{\mu\nu}^{(m)}F^{\mu\nu (n)}\psi_m\psi_n\right.\nn\\
&&~~~\left.
+\frac{9}{8}\frac{U_z^3}{\Ukk}
\left(
\del_\mu \varphi^{(m)}\del^\mu \varphi^{(n)}\phi_m\phi_n
+B_\mu^{(m)}B^{\mu(n)}\dot\psi_m\dot\psi_n
-2\del_\mu\varphi^{(m)}B^{\mu(n)}\phi_m\dot\psi_n
\right)
\right].\nn\\
\label{phiphiBB}
\end{eqnarray}

Let us first consider the vector meson field $B_\mu^{(m)}$.
Omitting $\varphi^{(m)}$, we obtain
\begin{eqnarray}
S_{D8}
=-\wt T(2\pi\alpha')^2
\int d^4x dz\, \sum_{m,n}
\left[\,
\frac{R^3}{4U_z}
F_{\mu\nu}^{(m)}F^{\mu\nu (n)}\psi_m\psi_n
+\frac{9}{8}\frac{U_z^3}{\Ukk}
B_\mu^{(m)}B^{\mu(n)}\dot\psi_m\dot\psi_n
\right].\nn\\
\label{Bonly}
\end{eqnarray}
It is useful to define the quantities
\begin{equation}
Z\equiv\frac{z}{\Ukk} \ ,
~~~ K(Z)\equiv 1+Z^2=\left(\frac{U_z}{\Ukk}\right)^3 \ .
\end{equation}
Using these, the above action can be written
\begin{eqnarray}
S_{D8}= -\wt T(2\pi\alpha')^2R^3
\int d^4x dZ \sum_{n,m}\bigg[
&&
\frac{1}{4}\,K^{-1/3}\,F_{\mu\nu}^{(n)}F^{(m)\mu\nu}\psi_n\psi_m
\nn \\ 
&&
+\frac{1}{2}\Mkk^2\,K\,B_{\mu}^{(n)}B^{(m)\mu}
\,\del_Z\psi_n\,\del_Z\psi_m \bigg] \ .
\end{eqnarray}
Now we choose $\psi_n$ ($n\ge 1$) as the eigenfunctions satisfying
\begin{equation}
-K^{1/3}\,\del_Z\left(
K\,\del_Z\psi_n\right)=
\lambda_n\psi_n \ ,
\label{deqn;psi}
\end{equation}
with the normalization condition of $\psi_n$ given by
\begin{equation}
\wt T(2\pi\alpha')^2R^3 \int dZ K^{-1/3}\, \psi_n\psi_m =\delta_{nm} \ .
\label{norm;psi}
\end{equation}
{}From these, we obtain
\begin{eqnarray}
\wt T(2\pi\alpha')^2R^3 \int dZ\,K\,\del_Z\psi_m\,\del_Z\psi_n
=\lambda_n\delta_{nm}
\label{lambda}
\end{eqnarray}
and
\begin{eqnarray}
S_{D8}=
\int d^4x\, \sum_{n=1}^\infty
\left[\,
\frac{1}{4}
F_{\mu\nu}^{(n)}F^{\mu\nu (n)}
+\frac{1}{2}m_n^2\, B_\mu^{(n)}B^{\mu(n)}
\right] \ ,
\end{eqnarray}
with $m_n^2\equiv\lambda_n\Mkk^2$ being non-zero and positive
for all $n\ge 1$.
Thus $B_\mu^{(n)}$ is a massive vector meson of
mass $m_n$.

Next, we include $\varphi^{(n)}$. 
In order to canonically normalize the kinetic term for
$\varphi^{(n)}$, we impose
the ortho-normal condition
\begin{eqnarray}
(\phi_m,\phi_n)\equiv
\frac{9}{4}\,\wt T(2\pi\alpha')^2\Ukk^3
\int dZ \,K\,\phi_m \phi_n  =\delta_{mn} \ ,
\end{eqnarray}
for the complete set $\{\phi_n\}$.
{}From (\ref{lambda}), it is seen that we can choose 
$\phi_n=m_n^{-1}\,\dot\psi_n$
($n\ge 1$).
Note that there also exists a function $\phi_0$ that is
orthogonal to $\dot\psi_n$ for all $n\ge 1$.
Actually, if we take $\phi_0= C/K$, we have
\begin{eqnarray}
(\phi_0,\phi_n)\propto
\int dZ\, \del_Z\psi_n
=0 \ . ~~~~(\mbox{for}~n\ge 1)
\label{phi0}
\end{eqnarray}
The normalization constant $C$ can be determined by
\begin{eqnarray}
1=(\phi_0,\phi_0)
=\frac{9}{4}\,\wt T(2\pi\alpha')^2\Ukk^3\,\pi\, C^2 \ .
\end{eqnarray}
Then (\ref{F}) becomes
\begin{eqnarray}
F_{\mu z}&=&\del_\mu\varphi^{(0)}\phi_0+
\sum_{n\ge 1}\left(
m_n^{-1}\del_\mu \varphi^{(n)}
-B^{(n)}_\mu\right)\dot\psi_n \ .
\end{eqnarray}
Here, $\del_\mu\varphi^{(n)}$ can be absorbed into $B_\mu^{(n)}$
through the gauge transformation
\begin{eqnarray}
B^{(n)}_\mu\ra B^{(n)}_\mu+m_n^{-1}\del_\mu \varphi^{(n)}\ .
\end{eqnarray}
Then the action (\ref{phiphiBB}) becomes
\begin{eqnarray}
S_{D8}=-
\int d^4x \,
\left[\,
\half\,\del_\mu\varphi^{(0)}\del^\mu\varphi^{(0)}
+ \sum_{n\ge 1}\left(
\frac{1}{4}
F_{\mu\nu}^{(n)}F^{\mu\nu (n)}
+\frac{1}{2} m_n^2\,B_\mu^{(n)}B^{\mu(n)}
\right)
\right].
\label{S0}
\end{eqnarray}
We interpret $\varphi^{(0)}$ as the pion field,
which is the Nambu-Goldstone boson
associated with the chiral symmetry breaking.
This interpretation will become clearer when we
generalize the analysis to the multi-flavor case
in \S \ref{multiNf}.
The parity of this mode is determined
in \S \ref{numerical} (see also \S \ref{CP}), and
it turns out that this field is a pseudo-scalar meson,
as expected.

Since we are considering the $N_f=1$ case here, the spontaneously
broken chiral symmetry is the axial $U(1)_A$ symmetry,
and the associated NG boson is actually the analog of the
$\eta'$ meson.
However, because $U(1)_A$ is anomalous, it can be a massless NG boson
only in the large $N_c$ limit.
The supergravity description of the $U(1)_A$ anomaly and
the source of the $\eta'$ meson mass are discussed
in \S \ref{etamass}.

\subsection{$A_z=0$ gauge}
\label{Az=0}

In the previous subsection, we implicitly assumed
that the gauge potential vanishes in the limit $z\ra\pm\infty$.
In order to obtain a normalizable four-dimensional action,
the field strength should vanish as $z\ra\pm\infty$,
and then we can always choose a gauge such that
the gauge potential vanishes asymptotically for large $|z|$.
Here we make a comment on another gauge choice,
the $A_z=0$ gauge, which is used in later sections.
Because the massless pseudo-scalar meson
$\varphi^{(0)}$ appears in $A_z$ as in (\ref{expand;A}),
it might be thought that the meson would be gauged away in
the $A_z=0$ gauge. However, this is not the case.
It is important to note that we cannot choose a gauge
that simultaneously satisfies both $A_z=0$ and $A_\mu\ra 0$
($z\ra\pm\infty$).
In changing to the $A_z=0$ gauge from the previous one, the massless
pseudo-scalar meson appears in the asymptotic behavior
of $A_\mu$.

In the expansion (\ref{expand;A}),
the $A_z=0$ gauge is realized through the gauge transformation
\begin{eqnarray}
A_M\ra A_M -\del_M \Lambda \ ,
\label{gaugetr}
\end{eqnarray}
with
\begin{eqnarray}
\Lambda(x^\mu,z)=\varphi^{(0)}(x^\mu)\,\psi_0(z)
+\sum_{n=1}^\infty
 \varphi^{(n)}(x^\mu)\,m_n^{-1}\psi_n(z) \ ,
\end{eqnarray}
where $\psi_0$ is defined as
\begin{eqnarray}
\psi_0(z)=\int^z_0 dz'\, \phi_0(z')
=C\,\Ukk\,\arctan\left(\frac{z}{\Ukk}\right).
\end{eqnarray}
Note that $\psi_0$ can be thought of as the zero mode of
the eigenequation (\ref{deqn;psi}), though it is not
normalizable.
The gauge transformation (\ref{gaugetr}) implies
\begin{eqnarray}
&&A_z(x^\mu,z)=0 \ ,\nn\\
&&A_\mu(x^\mu,z)=-\del_\mu\varphi^{(0)}(x^\mu)\psi_0(z)
+\sum_{n\ge 1}
\left(B_\mu^{(n)}(x^\mu)-m_n^{-1}\del_\mu\varphi^{(n)}(x^\mu)\right)
\psi_n(z) \ .
\label{expAA}
\end{eqnarray}
We can absorb $m_n^{-1}\del_\mu\varphi^{(n)}$ in the second term of
(\ref{expAA}) into $B_\mu^{(n)}$, and as a result we obtain
\begin{eqnarray}
A_\mu(x^\mu,z)=-\del_\mu\varphi^{(0)}(x^\mu)\psi_0(z)
+\sum_{n\ge 1} B_\mu^{(n)}(x^\mu) \psi_n(z) \ .
\label{expA}
\end{eqnarray}
Here, a $z$-independent pure gauge term could be added, but it would
not contribute to the action, as
it could be gauged away through a residual $z$ independent
gauge transformation.
The $\varphi^{(0)}$ dependence enters the gauge potential $A_\mu$
as the boundary conditions
\begin{eqnarray}
A_\mu(x^\mu,z)\ra \mp C\Ukk\frac{\pi}{2}\,
\del_\mu\varphi^{(0)}(x^\mu)~~~~~(\mbox{as}~~ z\ra\pm\infty)
\label{Abdry}
\end{eqnarray}
in the $A_z=0$ gauge. A general
gauge configuration with the boundary conditions
(\ref{Abdry}) can be expanded as (\ref{expA}).
Although $\psi_0$ in the expansion is non-normalizable,
the field strength is normalizable, and the action remains finite.
In fact, because the action is independent of the gauge
choice, we, of course, reproduce the action (\ref{S0}) by
inserting the expansion (\ref{expA}) into (\ref{Bonly}).

\section{Analysis of meson spectra}
\label{numerical}

In this section, we report the results of a numerical
computation through which we determined the spectra of
(pseudo-) scalar and (axial-) vector mesons by studying normalizable
fluctuations around the D8 probe configuration.
We also determine their parity and compare them with
the observed mesons.
The fluctuations of fermions on the probe brane are considered in
Appendix \ref{fermion}.

\subsection{Vector mesons}

As discussed in the previous subsection, we have to solve (\ref{deqn;psi})
with the normalization condition given by (\ref{norm;psi}).
It is easily seen that the normalization condition is satisfied if
$\psi_n$ behaves as 
$\psi_n(z)\sim \cO(z^a)$ with $~a<-1/6$ as $z\ra \pm \infty$.

First, we note that the asymptotic behavior of $\psi_n$
(i.e., in the limit $z\rightarrow\infty$) is
\begin{equation}
\psi_n(z)\sim \cO(1)~~ \mbox{or}~~ \cO(z^{-1})
~~~~(\mbox{for}~~z\ra\infty) \ .
\label{psiasym}
\end{equation}
We choose $\psi_n\sim z^{-1}$ in order to have a normalizable solution.
It is then convenient to work with the new wave function defined by
\begin{equation}
\psit_n(Z)\equiv Z \,\psi_n(\Ukk Z) \ ,
\end{equation}
which asymptotically behaves as
\begin{eqnarray}
\psit_n(Z)\sim \cO(1) \ .~~~~ (\mbox{for}~~Z\ra\infty)
\end{eqnarray}
In terms of $\psit_n$, (\ref{deqn;psi}) reads
\begin{equation}
K\del_Z^2\psit_n
-\frac{2}{Z}\del_Z\psit_n
+\left(\frac{2}{Z^2}+\lambda_n K^{-1/3}\right)
\psit_n=0 \ .
\end{equation}
Using the new variable $Z=e^{\eta}$, this can be recast as
\begin{equation}
\del_{\eta}^2\psit_n+A\del_{\eta}\psit_n+B\psit_n=0 \ ,
\end{equation}
where
\begin{eqnarray}
A&\!=\!&-\frac{1+3e^{-2\eta}}{1+e^{-2\eta}}
=\sum_{l=0}^\infty A_l\,e^{-\frac{2l}{3}\eta} \ ,
\nn \\
B&\!=\!&\frac{2e^{-2\eta}}{1+e^{-2\eta}}
+\lambda_ne^{-\frac{2}{3}\eta}(1+e^{-2\eta})^{-4/3}
=\sum_{l=0}^\infty B_l\,e^{-\frac{2l}{3}\eta} \ .
\end{eqnarray}
For instance, we have
\begin{eqnarray}
&&A_0=-1,~A_1=A_2=0,~A_3=-2,~A_4=A_5=0,\cdots\ , \nn \\
&&B_0=0,~B_1=\lambda_n,~B_2=0,~B_3=2,~B_4=-\frac{4}{3}\lambda_n,
~B_5=0,\cdots .
\end{eqnarray}
Then, by expanding $\psit_n$ as
\begin{equation}
\psit_n=\sum_l \alpha_l\, e^{-\frac{2l}{3}\eta} \ ,
\end{equation}
with $\alpha_0=1$, 
it is easy to verify that the quantities
 $\alpha_l$ obey the recursion relation
\begin{equation}
\frac{4l^2}{9}\alpha_l-\frac{2}{3}\sum_{m=1}^{l}mA_{l-m}\alpha_m
+\sum_{m=0}^{l-1}B_{l-m}\alpha_m=0 \ .
\end{equation}
This yields the following:
\begin{equation}
\alpha_1=-\frac{9}{10}\lambda_n \ , ~~
\alpha_2=\frac{81}{280}\lambda_n^2 \ , ~~
\alpha_3=-\frac{1}{3}-\frac{27}{560}\lambda_n^3 \ , \cdots .
\end{equation}
We used these data as an input to solve (\ref{deqn;psi}) numerically
by means of a shooting method.
In this computation, we can assume $\psi_n$ to be an
even or odd function since Eq. (\ref{deqn;psi})
is invariant under $Z\ra -Z$, and we impose
the regularity conditions
\begin{equation}
\partial_Z\psi_n(0)=0~~\mbox{or}~~\psi_n(0)=0 \ ,
\end{equation}
at $Z=0$ for even and odd functions $\psi_n$, respectively.

Our study produced the following result:
\begin{equation}
\lambda_n^{CP}=\ 0.67^{--} \ , ~~ 1.6^{++}\ , ~~ 2.9^{--}\ ,
 ~~ 4.5^{++}\ ,~\cdots \ .
\end{equation}
Here $C$ and $P$ stand for charge conjugation and parity.
To read off the parity, recall that
the action is invariant under
the transformation $(x^1,x^2,x^3,z)\ra (-x^1,-x^2,-x^3,-z)$,
which is an element of the five-dimensional
proper Lorentz transformation. This transformation is
interpreted as the parity transformation in the four-dimensional
theory. Then, from the expansion (\ref{expA}),
we see that
$B_{\mu}^{(n)}$ is a four-dimensional vector and axial vector 
when $\psi_n$ is an even and an odd function, respectively.
Regarding the charge conjugation property of $B_{\mu}^{(n)}$,
we show in \S \ref{CP}
 that $B_{\mu}^{(n)}$ is odd (even) when $\psi_n(Z)$
is an even (odd) function.
Because $\psi_{2k}$ is odd and $\psi_{2k+1}$ is even,
the lightest mode, $\lambda_1$, gives a vector meson with $C=-1$, and
we interpret it as the $\rho$ meson.
The second lightest one, $\lambda_2$, is an axial-vector meson with
$C=+1$, which is interpreted as the $a_1(1260)$ meson
(see, e.g., Ref.~\citen{pdg} for experimental data concerning mesons).
The third one, $\lambda_3$, has $C=P=-1$, and therefore is identified
with $\rho(1450)$.\footnote{The spin 1 meson spectrum with alternating
parity is observed in the open moose model \cite{SS}.}
Similarly, the massless meson $\varphi^{(0)}$ turns out to be
a pseudo-scalar meson, since $\psi_0$ is an odd function.
This is consistent with the interpretation given in
\S \ref{fluc}.

Although we know that our model deviates from QCD above
an energy scale around $\Mkk$,
it is tempting to compare our results
with the observed meson table \cite{pdg}.
The ratio of $\lambda_2$ to $\lambda_1$ should be compared
with the ratio of $m_{a_1(1260)}^2$ to $m_\rho^2$. The result is
\begin{eqnarray}
&&\frac{\lambda_2}{\lambda_1}\simeq
\frac{1.6}{0.67}
\simeq ~2.4~~~(\mbox{our model}) \ ,\nn\\
&&\frac{m_{a_1(1260)}^2}{m_\rho^2}
\simeq\frac{(1230~ \mbox{MeV})^2}{(776~ \mbox{MeV})^2}
\simeq ~2.51~~~(\mbox{experiment}) \nn \ .
\end{eqnarray}
Also, the ratio of the mass squared of the $\rho(1450)$ meson to that
of the $\rho$ meson is estimated as
\begin{eqnarray}
&&\frac{\lambda_3}{\lambda_1}\simeq
\frac{2.9}{0.67}
\simeq ~4.3~~~(\mbox{our model}) \ ,\nn\\
&&\frac{m_{\rho(1450)}^2}{m_\rho^2}
\simeq\frac{(1465~ \mbox{MeV})^2}{(776~ \mbox{MeV})^2}
\simeq ~3.56~~~(\mbox{experiment}) \nn\ .
\end{eqnarray}

\subsection{Massive scalar mesons}
\label{msm}
Here, we follow the same procedure as in the previous subsection
to obtain the meson spectrum from the fluctuation
of the scalar field $y$ on the D8-brane.
The action of $y$ can be read from (\ref{fluc;y}), which
is rewritten as
\begin{equation}
S_{D8}\simeq -\frac{4}{9}\,\wt T  R^3
\int d^4x dZ\left[\,
\frac{1}{2}K^{-1/3}\left(\del_{\mu} y\right)^2+\frac{\Mkk^2}{2}
\left( K(\del_Z y)^2+2 y^2\right)
\,\right] \ .
\end{equation}
We expand $y$ as
\begin{equation}
y(x^\mu,z)=\sum_{n=1}^\infty\cu^{(n)}(x^\mu)\rho_n(Z) \ ,
\label{expand;Y}
\end{equation}
where $\{\rho\}_{n\ge 1}$ is the complete set of the
eigenequation
\begin{equation}
K^{1/3}\bigg[
-\del_Z(K\del_Z\rho_n)+2\rho_n\bigg]=\lambda^{\prime}_n\rho_n \ ,
\label{deqn;rho}
\end{equation}
with the normalization condition given by
\begin{equation}
\frac{4}{9}\,\wt T R^3
\int dZ\,K^{-1/3}\,\rho_n\rho_m=\delta_{nm} \ .
\label{norm;rho}
\end{equation}
Then, the action becomes
\begin{equation}
S_{D8}=\frac{1}{2}\int d^4x\sum_n\left[
\left(\del_{\mu}\cu^{(n)}\right)^2
+\lambda^{\prime}_n\Mkk^2\left(\cu^{(n)}\right)^2
\right] \ .
\end{equation}
Thus, $\cu^{(n)}$ gives a scalar or pseudo-scalar meson with mass
squared given by $\lambda_n^{\prime}\Mkk^2$.
To understand their parity nature, we note that $y$ is a
scalar field on the D8-brane world-volume.
This is because the CS coupling on it, 
\begin{equation}
S_{\rm CS}\sim \int_{D8} F\wedge F\wedge C_5+\cdots \ , ~~~~~
C_5 
\sim y \,dx^0\wedge \cdots \wedge dx^3\wedge dz \ ,
\end{equation}
dictates that $y$ is parity even.
Hence,
$\cu^{(n)}$ is a scalar (pseudo-scalar) meson when $\rho_n(Z)$ is
an even (odd) function, as seen from the decomposition
(\ref{expand;Y}).
We see below that $\cu^{(n)}$ is even (odd) under charge
conjugation when $\rho_n$ is even (odd).

It can be seen that the asymptotic behavior of $\rho_n$ is given by
\begin{equation}
\rho_n(Z)\sim \cO(Z)~~ \mbox{or}~~ \cO(Z^{-2}) \ .
~~~(\mbox{for}~~Z\ra\infty)
\end{equation}
We consider the normalizable solutions and define
\begin{equation}
\rhot_n(Z)=Z^2 \rho_n(Z) \ ,
\end{equation}
which behaves as $\rhot_n(Z)\sim 1$ for $Z\ra\infty$.
In terms of $\rhot_n(Z)$, (\ref{deqn;rho}) becomes
\begin{equation}
K\del_Z^2\rhot_n
-2\left(1+\frac{2}{Z^2}\right)Z\del_Z\rhot_n
+\left(\frac{6}{Z^2}+\lambda_n^{\prime}K^{-1/3}\right)\rhot_n=0 \ .
\end{equation}
Then, using $Z=e^{\eta}$, we obtain
\begin{equation}
\del_{\eta}^2\rhot_n+C\del_{\eta}\rhot_n+D\rhot_n=0 \ ,
\end{equation}
where
\begin{eqnarray}
C&\!=\!&-\frac{3+5e^{-2\eta}}{1+e^{-2\eta}}
=\sum_{l=0}^\infty C_l\,e^{-\frac{2l}{3}\eta} \ ,
\nn \\
D&\!=\!&\frac{6e^{-2\eta}}{1+e^{-2\eta}}
+\lambda_n'e^{-\frac{2}{3}\eta}(1+e^{-2\eta})^{-4/3}
=\sum_{l=0}^\infty D_l\,e^{-\frac{2l}{3}\eta} \ .
\end{eqnarray}
Next, expanding $\rhot_n$ as
\begin{equation}
\rhot=\sum_{l=0}^\infty \beta_l\, e^{-\frac{2l}{3}\eta} \ ,
\label{asym;rho}
\end{equation}
with $\beta_0=1$, 
we find that $\beta_l$ obeys the recursion relation
\begin{equation}
\frac{4l^2}{9}\beta_l-\frac{2}{3}\sum_{m=1}^{l}mC_{l-m}\beta_m
+\sum_{m=0}^{l-1}D_{l-m}\beta_m=0 \ .
\end{equation}
It follows that
\begin{equation}
\beta_1=-\frac{9}{22}\lambda^{\prime}_n \ , ~~
\beta_2=\frac{81}{1144}\lambda_n^{\prime 2} \ , ~~
\beta_3=-\frac{3}{5}-\frac{81}{11440}\lambda_n^{\prime 3}
\ , ~\cdots .
\end{equation}
We solve the differential equation (\ref{deqn;rho}) using a shooting
method with the boundary conditions specified by (\ref{asym;rho}).
As above, the regularity conditions
\begin{equation}
\partial_Z\rho_n(0)=0 ~~{\rm or}~~\rho_n(0)=0
\end{equation}
must be satisfied,
because $\rho_n$ is either an even function or an odd function.
The result is
\begin{equation}
\lambda_n^{\prime CP}=3.3^{++} \ ,~~ 5.3^{--} \ ,~\cdots \ .
\end{equation}
Considering the quantum numbers, we find that
the lightest mode should
be identified with $a_0(1450)$.\footnote{In
the meson summary tables in Ref.~\citen{pdg}, the lightest scalar
meson with $C=1$ and isospin $I=1$
is given by $a_0(980)$. However, we do not identify
the lightest massive scalar meson $\cu^{(1)}$
with $a_0(980)$, since it is often regarded as
a meson-meson resonance or a four-quark state.
(See the ``Note on Non-$q\bar q$ Mesons'' at the end of the Meson
Listings in Ref.~\citen{pdg}.)}
The ratio of the mass squared of the $a_0(1450)$ meson
to that of the $\rho$ meson is estimated as
\begin{eqnarray}
&&\frac{\lambda_1^{\prime}}{\lambda_1}\simeq
\frac{3.3}{0.67}
\simeq ~4.9~~~(\mbox{our model}) \ ,\nn\\
&&\frac{m_{a_0(1450)}^2}{m_\rho^2}
\simeq\frac{(1474~ \mbox{MeV})^2}{(776~ \mbox{MeV})^2}
\simeq ~3.61~~~(\mbox{experiment}) \nn\ .
\end{eqnarray}

\section{Multi-flavor case}
\label{multiNf}
In this section, we 
generalize the previous analysis to the case of $N_f>1$ flavor QCD
by introducing $N_f$ probe D8-branes. 

\subsection{Fluctuation modes around multiple probe branes}
\label{fl}
{}From the holographic point of view,
the effective action of the mesons is
given by the non-Abelian Born-Infeld action (plus CS-term)
of the probe D8-brane,
whose leading terms are given by the non-Abelian generalization of
(\ref{FF}),
\begin{eqnarray}
S_{D8}&=&\wt T(2\pi\alpha')^2
\int d^4x dz\, 2\tr\left[\,
\frac{R^3}{4U_z}
\eta^{\mu\nu}\eta^{\rho\sigma} F_{\mu \rho}F_{\nu \sigma}
+\frac{9}{8}\frac{U_z^3}{\Ukk}
\eta^{\mu\nu} F_{\mu z}F_{\nu z}
\right] \ ,
\label{multiFF}
\end{eqnarray}
where $F_{MN}=\del_M A_N-\del_N A_M +[A_M,A_N]$
is the field strength
of the $U(N_f)$ gauge field $A_M$ ($M=0,1,2,3,z$) on the D8-brane.
\footnote{In this section, we treat the gauge field $A_M$
as anti-Hermitian matrices. }
\footnote{
The extra factor of 2 in front of the trace is due to our
normalization $\Tr T^a T^b=\half\delta_{ab}$ of the generators $T^a$.
In order to compensate for this factor, we redefine $\wt T$ as
$\wt T=\frac{1}{3}R^{3/2}\Ukk^{1/2} \, T V_{4}\,g_s^{-1}$
in this section.}
In order to obtain a finite four-dimensional action for the modes
localized around $z=0$, the field strength $F_{MN}$
should vanish at $z=\pm\infty$.
This implies that the gauge field $A_M$ must asymptotically take
a pure gauge configuration:
\begin{eqnarray}
A_M(x^\mu,z)\ra U_\pm^{-1}(x^\mu,z)\del_M U_\pm(x^\mu,z) \ .
~~~~~({\mbox{as}}~z\ra\pm\infty) 
\end{eqnarray}
Because $\pi_4(U(N_f))=0$,
\footnote{Here we consider the general $N_f>2$ case.
$N_f=2$ is an exception, because $\pi_4(U(2))=\bZ_2$.}
we can find a $U(N_f)$-valued function $U(x^\mu,z)$ such that
\begin{eqnarray}
U(x^\mu,z)\ra U_\pm(x^\mu,z)\ .~~~(\mbox{as}~z\ra\pm\infty)
\end{eqnarray}
Applying the gauge transformation
\begin{eqnarray}
A_M(x^M)\ra A_M^g(x^M)
\equiv g(x^M)A_M(x^M)g^{-1}(x^M)+g(x^M)\del_M g^{-1}(x^M)
\label{gauge}
\end{eqnarray}
with $g(x^\mu,z)=U(x^\mu,z)$, we can make the gauge potential
vanish asymptotically for large $|z|$;
\begin{eqnarray}
A_M(x^\mu,z)\ra 0  \ .
~~~~~(\mbox{as}~z\ra\pm\infty) 
\label{asym}
\end{eqnarray}
Working in this gauge,
we still have a gauge symmetry induced by the gauge function $g$
satisfying
\begin{eqnarray}
\del_M g\ra 0 \ .~~~(\mbox{as}~z\ra\pm\infty) 
\end{eqnarray}
We interpret this residual gauge symmetry with
$g_\pm\equiv \lim_{z\ra\pm\infty} g(x^\mu,z)$
as an element of the chiral symmetry
$(g_+,g_-)\in U(N_f)_L\times U(N_f)_R$ in QCD. 
This interpretation is consistent with our result
given in \S \ref{488} that the probe D8 consists of
smoothly connected D8 and \AD 8 branes, each of which is 
responsible for $U(N_f)_{L,R}$.

Recall that the pion field in the chiral Lagrangian
is usually written as
\begin{eqnarray}
e^{2i\pi(x^\mu)/f_\pi}\equiv U(x^\mu)~~ \in U(N_f)\ ,
\label{Upi}
\end{eqnarray}
which transforms as
\begin{eqnarray}
U(x^\mu)\ra g_+ U(x^\mu) g_-^{-1}
\end{eqnarray}
under the chiral symmetry $(g_+,g_-)\in U(N_f)_L\times U(N_f)_R$.
{}From the above interpretation of the chiral symmetry
in our holographic model, we consider that
\footnote{This form of the pion field is also considered in Ref. \citen{SS}.}
\begin{eqnarray}
U(x^\mu)=P \exp\left\{-
\int_{-\infty}^\infty dz'\, A_z(x^\mu,z')
\right\}
\label{pion}
\end{eqnarray}
behaves as the pion field.
It is also useful to define
\begin{eqnarray}
\xi_\pm^{-1}(x^\mu)=P \exp\left\{-
\int_{0}^{\pm \infty} dz'\, A_z(x^\mu,z')\right\} .~~
\end{eqnarray}
The pion field (\ref{pion}) can then be written as
\begin{eqnarray}
U(x^\mu)=\xi_+^{-1}(x^\mu)\xi_-(x^\mu) \ .
\label{xixi}
\end{eqnarray}
Under the residual
gauge transformation, $\xi_\pm(x^\mu)$ transforms as
\begin{equation}
\xi_{\pm}(x^{\mu})\ra h(x^{\mu})\, \xi_{\pm}(x^{\mu})\,
g^{-1}_\pm \ ,
\label{hxig}
\end{equation}
where $h(x^\mu)=g(x^\mu,z=0)$ is the local gauge symmetry
at $z=0$. As we discuss in \S \ref{hls}, these
are the basic ingredients in the hidden local symmetry approach.

In order to follow the procedure given in \S \ref{Az=0},
we change to the $A_z=0$ gauge by 
applying the gauge transformation with the gauge function
\begin{eqnarray}
g^{-1}(x^\mu,z)= P \exp\left\{-
\int_{0}^z dz'\, A_z(x^\mu,z')
\right\} \ ,
\label{ginv}
\end{eqnarray}
which changes the boundary conditions (\ref{asym})
for $A_\mu$ to
\begin{eqnarray}
A_\mu(x^\mu,z)\ra \xi_\pm(x^\mu)\del_\mu \xi^{-1}_\pm(x^\mu) \ .
~~~(\mbox{as}~z\ra \pm\infty) 
\label{asym2}
\end{eqnarray}
We can then expand the gauge field as in (\ref{expA}): 
\begin{eqnarray}
A_\mu(x^\mu,z)=\xi_+(x^\mu)\del_\mu \xi_+^{-1}(x^\mu)\psi_+(z)
+\xi_-(x^\mu)\del_\mu \xi_-^{-1}(x^\mu)\psi_-(z)
+\sum_{n\ge 1} B_\mu^{(n)}(x^\mu) \psi_n(z) \ ,\nn\\
\label{nonAexp0}
\end{eqnarray}
where $\psi_{\pm}$ is a zero mode of (\ref{deqn;psi}) with the
appropriate boundary conditions prescribed.
More explicitly, we have $\psi_\pm(z)\equiv \half\pm \wh\psi_0(z)$, with
\begin{eqnarray}
\wh\psi_0(z)&\equiv&\frac{1}{\pi}
\arctan\left(\frac{z}{\Ukk}\right) \ ,\\
\psi_\pm(z)&\equiv&\half\pm \frac{1}{\pi}
\arctan\left(\frac{z}{\Ukk}\right) \ .
\end{eqnarray}

The residual gauge symmetry which maintains $A_z=0$ is given by 
the following $z$-independent gauge transformation:
\begin{eqnarray}
A_\mu(x^\mu,z)\ra h(x^\mu) A_\mu(x^\mu,z) h^{-1}(x^\mu)+
h(x^\mu)\del_\mu h^{-1}(x^\mu) \ .
\label{A;underH}
\end{eqnarray}
The component fields in the expansion (\ref{nonAexp0}) are then
transformed as 
\begin{eqnarray}
B_\mu^{(n)}(x^\mu)\ra h(x^\mu) B_\mu^{(n)}(x^\mu) h^{-1}(x^\mu) \ ,
\label{homo}
\end{eqnarray}
along with (\ref{hxig}) for $\xi_\pm(x^\mu)$.

Because the parity transformation interchanges $\xi_+$ and $\xi_-$,
it is also convenient to rewrite (\ref{nonAexp0}) in parity
eigenmodes as
\begin{eqnarray}
A_\mu(x^\mu,z)
&\!=\!&
\left(
\xi_+(x^\mu)\del_\mu \xi_+^{-1}(x^\mu)-\xi_-(x^\mu)\del_\mu
\xi_-^{-1}(x^\mu)\right)\wh\psi_0(z)+\nn\\
&&+\half\left(
\xi_+(x^\mu)\del_\mu \xi_+^{-1}(x^\mu)+\xi_-(x^\mu)\del_\mu
\xi_-^{-1}(x^\mu)\right)
+\sum_{n\ge 1} B_\mu^{(n)}(x^\mu) \psi_n(z)\nn\\
&\!\equiv\!&\alpha_\mu(x^\mu)\wh\psi_0(z)+\beta_\mu(x^\mu)+
\sum_{n\ge 1} B_\mu^{(n)}(x^\mu) \psi_n(z) \ ,
\label{alphabeta}
\end{eqnarray}
where
\begin{eqnarray}
\alpha_\mu(x^\mu)&\equiv&
\xi_+(x^\mu)\del_\mu \xi_+^{-1}(x^\mu)
-\xi_-(x^\mu)\del_\mu \xi_-^{-1}(x^\mu) \nn\\
&=&\xi_-(x^\mu) \left(U^{-1}(x^\mu) \del_\mu U(x^\mu) \right)
\xi_-^{-1}(x^\mu) \ ,\label{alpha}\\
\beta_\mu(x^\mu)&\equiv&\half\left(
\xi_+(x^\mu)\del_\mu \xi_+^{-1}(x^\mu)
+\xi_-(x^\mu)\del_\mu \xi_-^{-1}(x^\mu)\right) \ .
\label{beta}
\end{eqnarray}
Note that the residual gauge symmetry (\ref{hxig})
acts on $\alpha_{\mu}$ and $\beta_{\mu}$ as
\begin{equation}
\alpha_{\mu}\ra h\,\alpha_{\mu}\,h^{-1} \ ,~~~
\beta_{\mu}\ra h\,\beta_{\mu} h^{-1}+h \del_{\mu} h^{-1} \ .
\end{equation}

In the following subsections,
we often employ a gauge such that $\xi_-(x^\mu)=1$ and
$U(x^\mu)=\xi_+^{-1}(x^\mu)$, in which the expansion (\ref{nonAexp0})
becomes
\begin{eqnarray}
A_\mu(x^\mu,z)=U^{-1}(x^\mu)\del_\mu U(x^\mu)\psi_+(z)
+\sum_{n\ge 1} B_\mu^{(n)}(x^\mu) \psi_n(z) \ .
\label{nonAexp}
\end{eqnarray}
Another convenient gauge choice is that with $\xi_+^{-1}(x^\mu)=
\xi_-(x^\mu)\equiv\xi(x^\mu)=\exp(i\pi(x^\mu)/f_\pi)$.
In this gauge, we find
\begin{eqnarray}
\alpha_\mu&=&\left\{\xi^{-1},\del_\mu\xi\right\}=
\frac{2i}{f_\pi}\del_\mu\pi+\cO(\pi^3) \ , \nn \\
\beta_\mu&=&\half\left[\xi^{-1},\del_\mu\xi\right]=
\frac{1}{2f_\pi^2}\left[\pi,\del_\mu\pi\right]+\cO(\pi^4) \ .
\label{ab}
\end{eqnarray}

\subsection{Pion effective action}
\label{pea}
Let us first omit the vector meson fields $B^{(n)}_\mu$
($n\ge 1$) and consider the effective action for the pion field
$U(x^\mu)$. Here we use the expansion (\ref{nonAexp}).
Then the field strength is
\begin{eqnarray}
F_{\mu\nu}&=&\left[U^{-1}\del_\mu U,U^{-1}\del_\nu U\right]\,
\psi_+(\psi_+-1) \ , \nn\\
F_{z \mu}&=&U^{-1}\del_\mu U\, \wh\phi_0 \ ,
\end{eqnarray}
where
\begin{eqnarray}
\wh\phi_0(z)\equiv\del_z\psi_+(z)=\frac{\Ukk^2}{\pi}\frac{1}{U_z^3(z)}\ .
\end{eqnarray}
Inserting this into the action (\ref{multiFF}),
we obtain
\begin{eqnarray}
S_{D8}&\!=\!&\wt T(2\pi\alpha')^2\int d^4 x dz
\,2\tr\left(
\frac{R^3}{4U_z}\psi_+^2(\psi_+-1)^2
\left[U^{-1}\del_\mu U,U^{-1}\del_\nu U\right]^2
+\frac{9}{8}\frac{U_z^3}{\Ukk}\wh\phi_0^2 (U^{-1}\del_\mu U)^2
\right)\nn\\
&\!=\!&\wt T(2\pi\alpha')^2\int d^4 x\,\tr\left(
A(U^{-1}\del_\mu U)^2+
B\left[U^{-1}\del_\mu U,U^{-1}\del_\nu U\right]^2
\right) \ ,
\label{pionEA}
\end{eqnarray}
where
\begin{eqnarray}
A&\equiv&2\int dz\,\frac{9}{8}\frac{U_z^3}{\Ukk}\wh\phi_0^2
=\frac{9\,\Ukk}{4\pi} \ ,\nn\\
B&\equiv&2\int dz\,\frac{R^3}{4U_z}\psi_+^2(\psi_+-1)^2
=\frac{R^3 b}{2\pi^4} \ .
\end{eqnarray}
Here $b$ is a numerical constant which is evaluated as
\begin{eqnarray}
b\equiv\int \frac{dZ}{(1+Z^2)^{1/3}}
\left(\arctan Z+\frac{\pi}{2}\right)^2
\left(\arctan Z-\frac{\pi}{2}\right)^2
\simeq 15.25\cdots \ .
\end{eqnarray}
The effective action (\ref{pionEA}) coincides with
that of the Skyrme model. Actually, the action of the Skyrme model is
(for a review see Ref.~\citen{skyrme;rev})
\begin{eqnarray}
S=\int d^4 x\left(\frac{f_\pi^2}{4}\tr\left(
 U^{-1}\del_\mu U\right)^2+\frac{1}{32 e^2}
\tr\left[U^{-1}\del_\mu U,U^{-1}\del_\nu U\right]^2
\right) \ ,
\label{Skyrme}
\end{eqnarray}
where $f_\pi\simeq 93$~MeV is the pion decay constant and $e$ is
a dimensionless parameter.
Comparing (\ref{pionEA}) with (\ref{Skyrme}), we obtain
\begin{eqnarray}
f_\pi^2=4\,\wt T(2\pi\alpha')^2A
=\frac{1}{54\pi^4}(\gym^2 N_c) \Mkk^2 N_c 
\label{fpi}
\end{eqnarray}
and
\begin{eqnarray}
e^2=\frac{1}{32\,\wt T(2\pi\alpha')^2 B}
=\frac{27\pi^7}{2 b}\frac{1}{(\gym^2N_c)N_c} \ .
\label{e}
\end{eqnarray}
Here we have used the useful relation
\begin{eqnarray}
\wt T(2\pi\alpha')^2=\frac{1}{108\pi^3}\Mkk N_c l_s^{-2} \ ,
\label{wtT}
\end{eqnarray}
as well as (\ref{RUg}).

Note that the $N_c$ dependence of the parameters
$f_\pi$ and $e$ in large $N_c$ limit with
fixed 't Hooft coupling $\gym^2 N_c$ can be read off of
(\ref{fpi}) and (\ref{e}) as $f_\pi \sim\cO(\sqrt{N_c})$
and $e\sim\cO(1/\sqrt{N_c})$, which 
are, of course, in agreement with the result obtained in large $N_c$ QCD.

\subsection{Including vector mesons}
\label{incl}
It is not difficult to include the vector meson fields $B^{(n)}_\mu$
in the above analysis.
Let us now include the lightest vector meson $B^{(1)}_\mu$, which is
identified as the $\rho$ meson, in the expansion
(\ref{alphabeta}). This gives
\begin{eqnarray}
A_\mu(x^\mu,z)=\alpha_\mu(x^\mu)\wh\psi_0(z)
+\beta_\mu(x^\mu)+ v_\mu(x^\mu) \psi_1(z) \ ,
\end{eqnarray}
where $v_\mu$ stands for $B^{(1)}_\mu$.
Then, using (\ref{ab}),
the field strength $F_{\mu\nu}$ is
\begin{eqnarray}
F_{\mu\nu}
&\!=\!&\frac{2i}{f_\pi}\left([\del_\mu\pi,v_\nu]+[v_\mu,\del_\nu\pi]
\right)\psi_1\wh\psi_0+\frac{1}{f_\pi^2}[\del_\mu\pi,\del_\nu\pi]
(1-4\wh\psi_0^2)
\nn\\
&&+(\del_\mu v_\nu-\del_\nu v_\mu)\psi_1+[v_\mu,v_\nu]\psi_1^2
+\cO((\pi,v_\mu)^3) \ .
\end{eqnarray}
Similarly, $F_{z\mu}$ is given by
\begin{eqnarray}
F_{z \mu}&\!=\!&\alpha_\mu\, \wh\phi_0+v_\mu\dot\psi_1 \nn\\
&\!=\!&\frac{2i}{f_\pi}\del_\mu\pi\,\wh\phi_0+v_\mu\dot\psi_1
+\cO(\pi^3)\ .
\end{eqnarray}
The effective action is obtained by inserting these into the action
(\ref{multiFF}).
For this purpose, we use the relation
\begin{eqnarray}
\int dz\, \tr\left[\,
\frac{R^3}{4U_z}F_{\mu \nu}^2\right]
&\!=\!&
\tr(\del_\mu v_\nu-\del_\nu v_\mu)^2
\int dz\,\frac{R^3}{4U_z}\psi_1^2
\nn\\&&
+2\tr\left(
[v_\mu,v_\nu](\del^\mu v^\nu-\del^\nu v^\mu)\right)
\int dz\,\frac{R^3}{4U_z}\psi_1^3\nn\\
&&+\frac{2}{f_\pi^2}
\tr\left([\del_\mu\pi,\del_\nu\pi](\del^\mu v^\nu-\del^\nu
v^\mu)\right)
\int dz\,\frac{R^3}{4U_z}\psi_1(1-4\wh\psi_0^2)+
\cO((\pi,v_{\mu})^4) \ ,
\nn\\ \\
\int dz\, \tr\left[\,
\frac{9}{8}\frac{U_z^3}{\Ukk}
F_{\mu z}^2\right]&\!=\!&\frac{-4}{f_\pi^2}
\tr(\del_\mu\pi\del^\mu\pi)
\int dz\, \frac{9}{8}\frac{U_z^3}{\Ukk}\wh\phi_0^2+
\tr v_\mu^2\int dz\, \frac{9}{8}\frac{U_z^3}{\Ukk}\dot\psi_1^2
+\cO((\pi,v_{\mu})^4) \ . 
\nn\\
\end{eqnarray}
Now, the effective action becomes
\begin{eqnarray}
S_{D8}&\!=\!&\int d^4 x \Bigg[
-a_{\pi^2} \tr\left(\del_\mu\pi\del^\mu\pi\right)
+a_{v^2}\left(
\frac{1}{2}\tr(\del_\mu v_\nu-\del_\nu v_\mu)^2
+m_v^2\tr v_\mu^2\right)
\nn\\
&&\frac{}{}~~~ +a_{v^3}\tr\Big(
[v_\mu,v_\nu](\del^\mu v^\nu-\del^\nu v^\mu)
\Big)+a_{v\pi^2}
\tr\Big([\del_\mu\pi,\del_\nu\pi](\del^\mu v^\nu-\del^\nu v^\mu)
\Big)\Bigg]\nn\\
&&~~~~+\cO((\pi,v_{\mu})^4)
\ .
\label{pirho}
\end{eqnarray}
Let us next determine the coefficients.
First,  we have
\begin{eqnarray}
a_{\pi^2}=
2\wt T(2\pi\alpha')^2\frac{4}{f_\pi^2}
\int dz\,\frac{9}{8}\frac{U_z^3}{\Ukk}\wh\phi_0^2=
\wt T(2\pi\alpha')^2\frac{9\Ukk}{\pi f_\pi^2}=1 \ ,
\end{eqnarray}
by definition (\ref{fpi}).
Also, it follows from (\ref{norm;psi}) that
\begin{eqnarray}
a_{v^2}=\wt T(2\pi\alpha')^2\int dz\, \frac{R^3}{U_z}\psi_1^2=1 \ .
\end{eqnarray}
The vector meson mass $m_v^2$ is
\begin{eqnarray}
m_v^2=
\wt T(2\pi\alpha')^2\int dz\,\frac{9}{4}\frac{U_z^3}{\Ukk}\dot\psi_1^2
=\lambda_1\Mkk^2 \ ,
\label{mv2}
\end{eqnarray}
as seen in \S \ref{fluc}.
The three-point couplings $a_{v^3}$ and $a_{v\pi^2}$ are
\begin{eqnarray}
a_{v^3}=
\wt T(2\pi\alpha')^2\int dz\,\frac{R^3}{U_z}\psi_1^3
=\frac{(6\pi)^{3/2}}{\sqrt{N_c(\gym^2 N_c)}}\,I_{v^3}\ ,
\label{av3}
\end{eqnarray}
and
\begin{eqnarray}
a_{v\pi^2}=
\frac{\wt T(2\pi\alpha')^2}{f_\pi^2}
\int dz\, \frac{R^3}{U_z}\psi_1(1-4\wh\psi_0^2)
=\frac{\pi(3\pi)^{3/2}}
{\Mkk^2\sqrt{2N_c(\gym^2 N_c)}}\,I_{v\pi^2} \ ,
\label{avpi2}
\end{eqnarray}
where $I_{v^3}$ and $I_{v\pi^2}$ are numerical constants
defined as
\begin{eqnarray}
I_{v^3}&=&\int dZ\,K^{1/6} \Psi_1^3 \ ,\\
I_{v\pi^2}&=&\int dZ\,\left(1-\frac{4}{\pi^2}\arctan^2(Z)\right)
K^{-1/6}\Psi_1 \ .
\end{eqnarray}
Here we have defined $\Psi_n(Z)$ as
\begin{eqnarray}
\Psi_n(Z)=\sqrt{\wt T(2\pi\alpha')^2R^3}~K(Z)^{-1/6}\psi_n(\Ukk Z)
\ ,
\end{eqnarray}
so that the normalization condition of $\Psi_n$
takes the form
\begin{eqnarray}
\int dZ\, \Psi_m(Z) \Psi_n(Z)=\delta_{mn} \ .
\end{eqnarray}
The numerical analysis described in \S \ref{numerical} yields
\begin{eqnarray}
\lambda_1\simeq 0.67 \ ,~~
I_{v^3}\simeq 0.45 \ ,~~
I_{v\pi^2}\simeq 1.6 \ .
\label{num}
\end{eqnarray}

\subsection{Comparison to the hidden local symmetry approach}
\label{hls}
In this section, we show that the D4/D8 model embodies the idea
of the hidden local symmetry to construct the effective action of the
pion and vector mesons.
To this end, we first give a brief review of
the hidden local symmetry approach
(for a comprehensive review, see Refs.~\citen{bky;review} and
\citen{Meissner}).

We write the pion field as (\ref{xixi}) and assume that
the action is invariant under the global chiral symmetry
group $G_{\rm global}=U(N_f)_L\times U(N_f)_R$ and a `hidden'
local symmetry group $H_{\rm local}= U(N_f)_V$,
which acts as (\ref{hxig}) for
$(g_+,g_-)\in G_{\rm global}$ and $h(x^\mu)\in H_{\rm local}$.
In addition, we introduce the gauge potential of $H_{\rm local}$,
denoted by $V_{\mu}$, which transforms as
\begin{eqnarray}
V_\mu(x^\mu)&\ra& h(x^\mu)V_\mu(x^\mu)h^{-1}(x^\mu)
+h(x^\mu)\del_\mu h^{-1}(x^\mu) \ 
\label{Vgauge}
\end{eqnarray}
under the local symmetry transformation $h(x^\mu)\in H_{\rm local}$.
Imposing the $G_{\rm global}\times H_{\rm local}$ symmetry
and parity, one can construct the following general Lagrangian
up to second order in derivatives of $\xi_\pm$:
\begin{eqnarray}
\cL_0&=&\cL_A+a\cL_V \ .
\label{HLaction0}
\end{eqnarray}
Here, $a$ is a constant and we have
\begin{eqnarray}
&&\cL_A=
\frac{f_\pi^2}{4}\tr(\alpha_{\mu})^2
=\frac{f_\pi^2}{4}\tr\, (U^{-1}\del_\mu U)^2 \ ,
\label{nlsm}\\
&&\cL_V=f_\pi^2\,\tr\left(V_\mu-\beta_{\mu}\right)^2 \ ,
\label{vbeta}
\end{eqnarray}
where $\alpha_\mu$ and $\beta_\mu$ are defined in
(\ref{alpha}) and (\ref{beta}).
The first term (\ref{nlsm}) is the Lagrangian density of
the usual non-linear sigma model associated with
$G/H$. At this stage, the second term (\ref{vbeta}) plays
no role. Because $V_\mu$ does not contain a kinetic term,
we can simply integrate it out.
Hence, the action defined as (\ref{HLaction0}) is equivalent
to that of the non-linear sigma model with arbitrary $a$.

A key step is to postulate that
the kinetic term for the gauge potential
$V_\mu$ emerges through a quantum effect
and that the total action is given by
\begin{eqnarray}
&&\cL^{\rm total}=\cL_A+a\cL_V+\cL_V^{\rm kin} \ ,
\label{HLaction}\\
&&\cL_V^{\rm kin}=\frac{1}{2g^2}\tr F^{V}_{\mu\nu} F^{V\,\mu\nu} \ ,~~~
F^{V}_{\mu\nu} = \del_\mu V_\nu-\del_\nu V_\mu+[V_\mu,V_\nu] \ .
\end{eqnarray}
Then the vector field $V_\mu$ becomes dynamical
and is identified as the $\rho$ meson.\cite{BKUYY,BKY,bky;review}
After rescaling as $V_\mu\ra gV_\mu$, we obtain
\begin{eqnarray}
\cL_V^{\rm kin}&=&\half\tr (\del_\mu V_\nu-\del_\nu V_\mu)^2
+g\tr\left( (\del_\mu V_\nu-\del_\nu V_\mu)[V^\mu,V^\nu]\right)+\cO(V^4)
\ ,\nn\\
\cL_A&=&-\tr \del_\mu\pi\del^\mu\pi+\cO(\pi^4) \ ,\nn\\
a\cL_V&=& a g^2 f_\pi^2\tr V_\mu^2
-ag\tr\left(V_\mu[\pi,\del^\mu\pi]\right)+\cO(\pi^4) \ .
\label{LV}
\end{eqnarray}
Then, $\cL_V$ can be written as
\begin{eqnarray}
a\cL_V&=& m_V^2\tr V_\mu^2
-2g_{V\pi\pi}\tr\left(V_\mu[\pi,\del^\mu\pi]\right)+\cO(\pi^4) \ ,
\label{LV2}
\end{eqnarray}
with the relations
\begin{eqnarray}
m_V^2=ag^2 f_\pi^2 \ ,~~~g_{V\pi\pi}=\frac{a}{2}g \ .
\label{mvgv}
\end{eqnarray}
The universality of the vector meson coupling
\begin{eqnarray}
g_{V\pi\pi}=g
\label{universal}
\end{eqnarray}
and the KSRF relation \cite{KS,RF}
\begin{eqnarray}
m_V^2=2g_{V\pi\pi}^2f_\pi^2
\label{KSRF}
\end{eqnarray}
hold when $a=2$.
\footnote{The experimental values are
\begin{eqnarray}
m_\rho\simeq 776~\mbox{MeV},~~f_\pi\simeq 92.6~\mbox{MeV},~~
g_{\rho\pi\pi}\simeq 5.99\,,
~~~~\frac{m_\rho^2}{g_{\rho\pi\pi}^2f_\pi^2}\simeq 1.96\ . \nn
\end{eqnarray}
}

The relation to the D4/D8 model is now clear.
As we saw in \S \ref{fl}, the model naturally possesses the
$G_{\rm global}\times H_{\rm local}$ symmetry as
part of the gauge symmetry on the probe D8-brane.
Furthermore, our model does contain the kinetic term of
the vector meson field, which results from the kinetic
term of the gauge potential on the D8-brane.

We next make a quantitative comparison between
the hidden local symmetry approach and our
model.
Note that the vector meson field $v_\mu$ appearing in \S \ref{incl}
transforms homogeneously under $H_{\rm local}$ as in (\ref{homo}),
while $V_\mu$ transforms as (\ref{Vgauge}).
Therefore, the gauge potential $V_\mu$ should be identified with
the vector field $v_\mu$ in the relation
\begin{eqnarray}
gV_\mu= gv_\mu+\beta_\mu \ .
\label{Vv}
\end{eqnarray}
Here $\alpha_\mu$ cannot enter, because it violates parity.
In order to compare the action (\ref{pirho}) with
(\ref{LV}), we rewrite the latter in terms
of $v_\mu$ and $\pi$ as
\begin{eqnarray}
\cL_V^{\rm kin}&=&\half\tr (\del_\mu v_\nu-\del_\nu v_\mu)^2
+g\tr\left( (\del_\mu v_\nu-\del_\nu v_\mu)[v^\mu,v^\nu]\right)\nn\\
&&~~~+\frac{1}{g f_\pi^2}\tr\left(
(\del_\mu v_\nu-\del_\nu v_\mu)[\del^\mu\pi,\del^\nu\pi]\right)
+\cO((\pi,v_\mu)^4) \ ,\\
a\cL_V&=& a g^2 f_\pi^2\tr v_\mu^2 \ .
\end{eqnarray}
Comparing this with (\ref{pirho}), we have
\begin{eqnarray}
m_v^2|_{\rm HLS}&=& a g^2 f_\pi^2=m_V^2 \ ,\\
a_{v^3}|_{\rm HLS}&=& g \ ,
\label{avvv}\\
a_{v\pi^2}|_{\rm HLS}&=&\frac{1}{g f_\pi^2}
=\frac{2g_{V\pi\pi}}{m_V^2} \ ,
\label{avpi2x}
\end{eqnarray}
where HLS represents the variables employed
in hidden local symmetry approach.
Note that (\ref{avvv}) and (\ref{avpi2x}) imply
$f_\pi^2 a_{v^3}a_{v\pi^2}|_{\rm HLS}=1$, while
Eqs. (\ref{fpi}), (\ref{av3}) and (\ref{avpi2}) imply
\begin{eqnarray}
f_\pi^2 a_{v^3}a_{v\pi^2}=I_{v^3} I_{v\pi^2}\ .
\label{faa}
\end{eqnarray}
According to the numerical analysis whose results are given in
(\ref{num}),
the right-hand side of (\ref{faa})
is approximately given by $I_{v^3}I_{v\pi^2}\simeq 0.72$, which
is fairly close to but disagrees with 1.
The disagreement cannot be resolved by adjusting the parameter $a$.
It is due to the fact that the action (\ref{HLaction}) is not
a general action with the assumed symmetry. In fact, one could add
terms like
\begin{eqnarray}
\tr\left(F_{\mu\nu}^V [\alpha^\mu,\alpha^\nu]\right),~~
\tr\left(F_{\mu\nu}^V (\del^\mu \beta^\nu-\del^\nu\beta^\mu
+[\beta^\mu,\beta^\nu])\right),~~
\tr\left(F_{\mu\nu}^V [V^\mu-\beta^\mu, V^\nu-\beta^\nu]\right)\nn\\
\end{eqnarray}
to the action (\ref{HLaction}), which contribute to the coefficients
$a_{v^3}$ and $a_{v\pi^2}$,
without breaking the $G_{\rm global}\times H_{\rm local}$ symmetry.
By contrast, the effective action for the D4/D8 model
has no such ambiguity, and even the higher derivative terms are 
in principle calculable in the framework of string theory.

To determine if the relations (\ref{universal}) and (\ref{KSRF})
are realized in our model, we use the relation
\begin{eqnarray}
g_{V\pi\pi}=\frac{m_v^2 a_{v\pi^2}}{2} \ ,
\end{eqnarray}
which is obtained by comparing the contribution to
the $\rho\ra\pi\pi$ decay width in the action (\ref{pirho})
with that in (\ref{LV2}).
Then the KSRF relation (\ref{KSRF}) reads
\begin{eqnarray}
m_v^2 a_{v\pi^2}^2 f_\pi^2=2 \ ,
\label{KSRF2}
\end{eqnarray}
for which the prediction of the hidden local symmetry approach is
\begin{eqnarray}
\left. m_v^2 a_{v\pi^2}^2 f_\pi^2\right|_{\rm HLS}=a \ .
\end{eqnarray}
In our model, the left-hand side of (\ref{KSRF2}) is
obtained from (\ref{fpi}), (\ref{mv2})  and (\ref{avpi2}) as
\begin{eqnarray}
m_v^2 a_{v\pi^2}^2 f_\pi^2=\frac{\pi}{4} I_{v\pi^2}^2\lambda_1\ ,
\label{KSRF3}
\end{eqnarray}
which is estimated as $\frac{\pi}{4}I_{v\pi^2}^2\lambda_1\simeq
1.3$ from (\ref{num}).

Similarly, the relation (\ref{universal}) can be written
\begin{eqnarray}
\frac{m_v^2 a_{v\pi^2}}{2 a_{v^3}}=1\ ,
\end{eqnarray}
while we obtain
\begin{eqnarray}
\frac{m_v^2 a_{v\pi^2}}{2 a_{v^3}}
=\frac{\pi}{8}\frac{I_{v\pi^2} \lambda_1}{I_{v^3}} \ 
\label{univ}
\end{eqnarray}
{}from (\ref{mv2})--(\ref{avpi2}). The numerical value
of the right-hand side is
$\frac{\pi}{8}\frac{I_{v\pi^2} \lambda_1}{I_{v^3}}\simeq 0.93$.

It is interesting that the combinations of the couplings
in (\ref{KSRF3}) and (\ref{univ})
are numerical constants that are uniquely 
fixed without any adjustable parameters in the holographic approach.

\subsection{Chiral anomaly and the WZW term}
\label{wzwterm}
In this subsection, we discuss the role of the CS-term
of the probe D8-brane. We argue that this term
yields the correct chiral anomaly of the dual QCD
as well as the Wess-Zumino-Witten term in the chiral Lagrangian.

The relevant term here is
\begin{eqnarray}
S^{D8}_{CS}&=&\mu \int_{D8} C_3 \tr F^3
\label{CS1}\\
&=&\mu \int_{D8} F_4\, \omega_5(A) \ ,
\label{CS2}
\end{eqnarray}
where $F_4=d C_3$ is the RR 4-form field strength and $\omega_5(A)$
is the Chern-Simons 5-form,
\begin{eqnarray}
\omega_5(A)=\tr\left(
AF^2-\half A^3 F+\frac{1}{10}A^5
\right) \ ,
\end{eqnarray}
which satisfies $d\omega_5=\tr F^3$.
The normalization constant is $\mu=1/48\pi^3$.
(See Appendix \ref{AppCS} for details.)
The equality between (\ref{CS1}) and (\ref{CS2})
holds only when the $F_4$ flux is an exact 4-form
and the surface term drops out.
In the case that there is a non-trivial $F_4$ flux, as in the
D4 background (\ref{D4sol}), we should use the expression
(\ref{CS2}) rather than (\ref{CS1}) \cite{Green-Harvey-Moore},
otherwise the CS-term (\ref{CS1}) vanishes, since we only include
the five-dimensional components of the gauge field 
$A_\mu,A_z$ ($\mu=0,1,2,3$) and assume that
they do not depend on the coordinates of the $S^4$.

The $F_4$ flux is associated with the D4-brane charge.
Integrating it over the $S^4$ in the D4 background (\ref{D4sol}),
we obtain
\begin{eqnarray}
\frac{1}{2\pi}\int_{S^4} F_4=N_c \ ,
\end{eqnarray}
which implies
\begin{eqnarray}
S^{D8}_{CS}=\frac{N_c}{24\pi^2}\, \int_{M^4\times \bR} \omega_5(A) \ ,
\label{D8CS}
\end{eqnarray}
where $M^4\times \bR$ is the five-dimensional plane parameterized by
 $x^0,\cdots,x^3$ and $z$.
Note, however, that there is an ambiguity in the expression (\ref{D8CS}),
since the CS 5-form $\omega_5(A)$ is not gauge invariant.
We postulate that the CS-term (\ref{D8CS}) is appropriate for
the gauge in which the gauge potential vanishes asymptotically 
as $z\ra\pm\infty$, as in (\ref{asym}).

In order to incorporate a background gauge field for the chiral
$U(N_f)_L\times  U(N_f)_R$ symmetry into our holographic model,
we utilize the non-normalizable mode in (\ref{psiasym})
and interpret the asymptotic value of the gauge potential
\footnote{This manner of incorporating background gauge potentials 
is in accord with that in the open moose model \cite{SS}.}
\begin{eqnarray}
A_{L\mu}(x^\mu)\equiv \lim_{z\ra+\infty} A_\mu(x^\mu,z),~~~
A_{R\mu}(x^\mu)\equiv \lim_{z\ra-\infty} A_\mu(x^\mu,z)
\end{eqnarray}
as the background gauge potential for the
$U(N_f)_L\times  U(N_f)_R$ symmetry.
Then, the infinitesimal gauge transformation on the D8-brane
($\delta_\Lambda A=d\Lambda+[A,\Lambda]$) gives
\begin{eqnarray}
\delta_\Lambda\omega_5(A)=d\omega_4^1(\Lambda,A) \ ,
\end{eqnarray}
where
\begin{eqnarray}
\omega_4^1(\Lambda,A)=\tr\left(
\Lambda\, d\left(AdA+\half A^3
\right)\right) \ ,
\end{eqnarray}
and hence the gauge transformation of the CS-term (\ref{D8CS}) is
\begin{eqnarray}
\delta_\Lambda S_{CS}^{D8}&=&\frac{N_c}{24\pi^2}\int_{M^4\times\bR}
d\omega_4^1(\Lambda,A)
=\frac{N_c}{24\pi^2}\int_{M^4}\left(
\omega_4^1(\Lambda,A)|_{z=+\infty}-\omega_4^1(\Lambda,A)|_{z=-\infty}
\right) \nn\\
&=&\frac{N_c}{24\pi^2}\int_{M^4}\left(
\omega_4^1(\Lambda_L,A_L)-\omega_4^1(\Lambda_R,A_R)
\right).
\label{anom}
\end{eqnarray}
This reproduces the chiral anomaly in QCD.

It is also convenient to work in the $A_z=0$ gauge.
We carry out the gauge transformation with the gauge
function given in (\ref{ginv}) to change to the $A_z=0$ gauge.
The gauge transformation of the CS 5-form $\omega_5(A)$ is given by
\begin{eqnarray}
\omega_5(A^{g})=\omega_5(A)+\frac{1}{10}\tr(gdg^{-1})^5
+d \alpha_4(dg^{-1} g,A) \ ,
\end{eqnarray}
with
\begin{eqnarray}
\alpha_4(V,A)=-\half\tr\left(
V(AdA+dAA+A^3)-\half VAVA-V^3 A
\right) \ ,
\end{eqnarray}
where $A^{g}=gA g^{-1}+gdg^{-1}$.
Then the CS-term (\ref{D8CS}) is obtained as
\begin{eqnarray}
S^{D8}_{CS}&=&
-\frac{N_c}{24\pi^2}\int_{M^4}\left(\alpha_4(d\xi_+^{-1}\xi_+,A_L)-
\alpha_4(d\xi_-^{-1} \xi_-,A_R)
\right)
\nn\\&&~~~~~+
\frac{N_c}{24\pi^2}\int_{M^4\times \bR}\left(\omega_5(A^{g})
-\frac{1}{10}\tr(gdg^{-1})^5\right)\ .
\label{CSWZW}
\end{eqnarray}
To see that this is equivalent to the WZW term appearing in the literature,
we expand $A^g$ as
\begin{eqnarray}
A^g_\mu(x^\mu,z)=
A_{L\mu}^{\xi_+}(x^\mu)\psi_+(z)+A_{R\mu}^{\xi_-}(x^\mu)\psi_-(z)
+\sum_{n\ge 1} B_\mu^{(n)}(x^\mu)\psi_n(z)
\label{bgexp}
\end{eqnarray}
and take the gauge such that $\xi_-(x^\mu)=1$
and $\xi_+^{-1}(x^\mu)=U(x^\mu)$.
Omitting $B^{(n)}_\mu$ for simplicity, we obtain
\begin{eqnarray}
\int_{M^4\times\bR}\omega_5(A^g)
&=&\half\int_{M^4}\tr\Bigg[
(A_L^{U^{-1}}A_R-A_RA_L^{U^{-1}}) d(A_L^{U^{-1}}+A_R)\nn\\
&&~~~~+\half A_L^{U^{-1}}A_RA_L^{U^{-1}}A_R
+\left((A_L^{U^{-1}})^3A_R-A_R^3A_L^{U^{-1}}\right)
\Bigg] \ .
\end{eqnarray}
After somewhat lengthy but straightforward algebra, we obtain
\begin{eqnarray}
S_{CS}^{D8}=-\frac{N_c}{48\pi^2}\int_{M^4}Z
-\frac{N_c}{240\pi^2}\int_{M^4\times \bR}\tr(gdg^{-1})^5 \ ,
\end{eqnarray}
where $g$ satisfies the boundary conditions
\begin{eqnarray}
\lim_{z\ra -\infty} g(x^\mu,z) =1 \ ,~~~~
\lim_{z\ra +\infty} g(x^\mu,z) =U^{-1}(x^\mu) \ ,
\end{eqnarray}
and $Z$ reads
\begin{eqnarray}
Z&=&
\tr[(A_R dA_R+dA_R A_R+A_R^3)(U^{-1}A_LU+U^{-1}dU)-{\rm{p.c.}}]\nn\\
&&+\tr[ dA_RdU^{-1}A_L U-{\rm{p.c.}}]
+\tr[A_R(dU^{-1}U)^3-{\rm{p.c.}}]\nn\\
&&+\half\tr[(A_RdU^{-1}U)^2-{\rm{p.c.}}]
+\tr[UA_R U^{-1}A_L dUdU^{-1}-{\rm{p.c.}}]\nn\\
&&-\tr[A_R dU^{-1}UA_R U^{-1}A_LU-{\rm{p.c.}}]
+\half\tr[(A_RU^{-1}A_LU)^2]\ .
\end{eqnarray}
Here ``p.c.'' represents the terms obtained by
exchanging $A_L\lra A_R,~U\lra U^{-1}$.
This is identical to the WZW term
in Refs.~\citen{Witten;WZW} and \citen{KRS;WZW}.

Using (\ref{CSWZW}) and (\ref{bgexp}), the couplings to the vector
mesons are also rather easy to work out. This will be explored elsewhere.

\subsection{Parity and charge conjugation}
\label{CP}
Recall that both parity and charge conjugation in QCD
interchange the left- and right-handed components of
the Dirac spinor. In terms of the D4/D8/\AD8 system
discussed in \S \ref{488}, this implies that parity
and charge conjugation interchange D8 and \AD8.
In our holographic description, interchanging D8 and \AD8
corresponds to the transformation $z\ra -z$ on the probe D8-brane,
though flipping the sign of $z$ alone does not keep the
CS-term (\ref{D8CS}) invariant.

As argued in \S \ref{numerical}, the parity transformation is
given by the transformation $(x^1,x^2,x^3,z)\ra (-x^1,-x^2,-x^3,-z)$.
Then, it acts on the component fields in (\ref{nonAexp0}) as
\begin{eqnarray}
\xi_\pm\ra \xi_\mp,~~U\ra U^{-1},~~\pi\ra -\pi,~~
B^{(n)}_\mu\ra (-1)^{n+1}B^{(n)}_\mu,
\end{eqnarray}
together with the coordinate 
transformation $(x^1,x^2,x^3,z)\ra (-x^1,-x^2,-x^3,-z)$.
Charge conjugation involves flipping the orientation of the string,
which amounts to taking the transpose of the gauge field on the probe D8-brane.
As can be easily shown, the transformation
$A\ra -A^T$ induces $F\ra -F^T$ and the CS 5-form transforms
$\omega_5(A)\ra\omega_5(-A^T)=-\omega_5(A)$. Therefore,
the transformation $A\ra -A^T$ together with $z\ra-z$ keeps
the action invariant. In terms of the component fields, it acts as
\begin{eqnarray}
\xi_\pm\ra\xi_\mp^*,~~U\ra U^{T},~~\pi\ra\pi^T,~~
B^{(n)}_\mu\ra (-1)^{n}B^{(n)T}_\mu,
\end{eqnarray}
while $z\ra -z$.

In summary, the massless meson $\pi$ has $J^{PC}=0^{-+}$, and 
the vector meson $B_\mu^{(n)}$ has $J^{PC}=1^{--}$ and
$J^{PC}=1^{++}$ for odd and even $n$, respectively.

The C-parity of the massive scalar mesons considered in \S \ref{msm}
is the same as the parity, since the scalar field $y$ is even under
charge conjugation for the same reason that it is even under parity.
Hence the field $\cu^{(n)}$ in the expansion (\ref{expand;Y})
represents a massive scalar meson with
$J^{PC}=0^{++}$ and $J^{PC}=0^{--}$ for
odd and even $n$, respectively.

\subsection{Baryon}
\label{baryon}
A baryon in the AdS/CFT context is 
realized as a D-brane wrapped around a sphere \cite{Witten;baryon}.
In our case, it is a D4-brane wrapped around the $S^4$.
By contrast, a baryon in the Skyrme model is realized
as a soliton  \cite{Skyrme:1,Skyrme:2,Skyrme:3}.
As we saw in \S \ref{pea}, the low-energy effective
theory on the D8-brane includes the Skyrme model, and hence
it is natural to expect that the Skyrmion and the wrapped
D4-brane are related. 
Actually, the wrapped D4-brane can be realized as a gauge configuration
carrying a non-trivial topological charge on the D8-brane.
Here we will explain that this topological charge is related
to the baryon number charge
and the Skyrmion constructed on the
D8-brane does correspond to the wrapped D4-brane.
\footnote{The relation between instantons in a five-dimensional gauge theory
and the Skyrmion was first pointed out in Ref.~\citen{SS}.}

Let us consider a static configuration of the gauge field on the D8-brane
and denote by $B\simeq \bR^4$ the four-dimensional space parameterized
by $(x^1,x^2,x^3,z)$.
The charge $n$ of the wrapped D4-brane is related to
the instanton number on $B$ as \cite{Douglas}
\begin{eqnarray}
\frac{1}{8\pi^2}\int_B \tr F^2=n \ .
\label{inst}
\end{eqnarray}
To see that the instanton number $n$ can be interpreted as the baryon
number, recall that the baryon number charge is defined as
$1/N_c$ times the charge of the diagonal $U(1)_V$ subgroup of
the $U(N_f)_V$ symmetry.
Inserting the gauge field $A=A_{\rm cl}+a\, 1_{N_f}$
on the probe D8-brane,
where $A_{\rm cl}$ is an instanton configuration satisfying
(\ref{inst}) and $a$ is a fluctuation of the $U(1)_V$ gauge field,
into the CS-term (\ref{D8CS}), we obtain
\begin{eqnarray}
S_{CS}^{D8}\simeq\frac{N_c}{8\pi^2}\int_{\bR\times B}
a\tr F^2_{\rm cl}\simeq
n N_c \int_\bR a\ ,
\label{nNa}
\end{eqnarray}
up to linear order with respect to the fluctuation $a$.
Here we have assumed that the instanton solution
possesses a point-like configuration.
The coupling (\ref{nNa}) implies that
the instanton configuration represents a point-like particle
with $U(1)_V$ charge $nN_c$;
that is, it is a particle of baryon number $n$.

Furthermore, using the relation
\begin{eqnarray}
\tr F^2 = d\omega_3(A),~~~\omega_3(A)=
\tr\left(AF-\frac{1}{3}A^3\right)
\end{eqnarray}
and the boundary conditions for the gauge field as in (\ref{nonAexp}),
\footnote{Here, because $\pi_3(U(N_f))$ is non-trivial, we cannot
employ the treatment given around (\ref{asym})
in order to make $A_\mu$ vanish as $z\ra \pm \infty$
through a gauge transformation within the static configuration.}
\begin{eqnarray}
 \lim_{z\ra+\infty} A_\mu(x^\mu,z)= U^{-1}(x^i)\del_i U(x^i)\ ,~~~
 \lim_{z\ra-\infty} A_\mu(x^\mu,z)=0\ ,
\end{eqnarray}
the baryon number can be expressed as
\begin{eqnarray}
n=\frac{1}{8\pi^2}\int_B \tr F^2
=\frac{1}{8\pi^2}\int_{\del B} \omega_3(A) \Big|_{z=\infty}
=-\frac{1}{24\pi^2}\int_{\bR^3} \tr (U^{-1}d U)^3 \ .
\end{eqnarray}
The last expression here 
counts the winding number of $U$, which represents
the homotopy group $\pi_3(U(N_f))\simeq\bZ$. It coincides
with the baryon number charge in
the Skyrme model \cite{Witten;WZW,Witten;b}.

The mass of the baryon is roughly approximated as the energy carried by
the D4-brane wrapped around the $S^4$, which can be read from the D4-brane
world-volume action,
\begin{eqnarray}
S_{D4}&=&
-\frac{1}{(2\pi)^4\, l_s^5 \,g_s}
\int_{\bR\times S^4} dx^0\,\epsilon_4
\sqrt{-g_{00}\,g_{(S^4)}}\, e^{-\phi}\nn\\
&=&-\frac{V_4}{(2\pi)^4\, l_s^5 \,g_s}
\left(\frac{\Ukk}{R}\right)^{3/4}(R^{3/2}\Ukk^{1/2})^2
 \left(\frac{R}{\Ukk}\right)^{3/4}
\int_{\bR} dx^0\nn\\
&=&-\frac{1}{27\pi}\Mkk (\gym^2 N_c)\,N_c\int_{\bR} dx^0 \ .
\end{eqnarray}
Thus the baryon mass is
\begin{eqnarray}
m_{\rm baryon}=\frac{1}{27\pi}\Mkk (\gym^2 N_c)\,N_c \ .
\end{eqnarray}
Note that we correctly obtain $m_{\rm baryon}\sim \cO(N_c)$,
as expected in large $N_c$ QCD.
To be more precise, we should solve the equations of motion
for the effective action of the D8-brane and work out
the energy carried by the solution, as done
in Ref.~\citen{anw;skyrme} for the Skyrme model.

\subsection{Axial $U(1)_A$ anomaly and the $\eta'$ mass}
\label{etamass}

In massless QCD, the axial $U(1)_A$ symmetry is broken
due to an anomaly.
However, in the large $N_c$ limit, the broken symmetry is
restored, because the anomaly is a subleading effect in the $1/N_c$
expansions. This means that in (and only in) the large $N_c$ limit, 
the spontaneous breaking of the $U(1)_A$ symmetry yields
a massless NG boson. This massless boson is regarded as
the $\eta'$ meson.
{}For a large but finite $N_c$, the $\eta'$ meson becomes massive,
because of the anomaly, gaining a mass that is $\cO(N_c^{-1})$.
In this subsection,
we discuss how the axial $U(1)_A$ anomaly and the mass
of the $\eta'$ meson can be understood in the supergravity
description.
A key observation was made in Ref.~\citen{ScWi}.
In that work, the anomaly cancellation of
the type IIB D9-\AD9 system is examined, and
it is shown that the gauge field corresponding to the $U(1)_A$
symmetry of the D9-\AD9 system
becomes massive by absorbing the RR 0-form field $C_0$.
As seen below, the $\eta'$ meson acquires mass
through an analogous mechanism in our brane configuration.
For previous closely related works on the $U(1)_A$ anomaly and
the $\eta'$ mass by means of supergravity, see Refs.~\citen{Barbon}
and \citen{Armoni} .

The axial $U(1)_A$ symmetry transformation in QCD acts on the quark fields $q^f$
($f=1,\cdots,N_f$) as
\begin{eqnarray}
q^f\ra e^{i\gamma_5 \alpha} q^f \ .
\end{eqnarray}
It is well known that the anomalous $U(1)_A$ transformation
is compensated for by the following shift of the theta parameter:
\begin{eqnarray}
\theta\ra\theta+ 2 N_f\alpha \ .
\end{eqnarray}
This relation can be seen in our supergravity description
as follows.
The theta parameter in QCD is related to the RR 1-form $C_1$
or its field strength $F_2$ as
\footnote{See Appendix \ref{AppCS} for our normalization
of the RR fields.}
\begin{eqnarray}
\theta = \int_{S^1} C_1=\int_{\bR^2} F_2 \ ,
\label{theta}
\end{eqnarray}
where $S^1$ is the circle parameterized by $\tau$
evaluated in the limit $U\ra\infty$ and $\bR^2$ is the two-plane
parameterized by $(U,\tau)$ or $(y,z)$ \cite{Witten;theta}.

When there exist $N_f$ D8-branes,
the RR fields are not invariant under the gauge transformation.
It is shown in Ref.~\citen{Green-Harvey-Moore}
that the anomaly cancellation requires that the RR 1-form
transform as
\begin{eqnarray}
\delta_\Lambda C_1=-i\,\tr(\Lambda)\delta(y) dy
\end{eqnarray}
under the infinitesimal $U(N_f)$ gauge transformation
$\delta_\Lambda A=d\Lambda+[\Lambda,A]$.
Then it follows from (\ref{theta}) that
the shift of the theta parameter is given by
\begin{eqnarray}
\delta_\Lambda\theta= -i\tr\Lambda|_{z=+\infty}
+i\tr\Lambda|_{z=-\infty} \ .
\end{eqnarray}
Recall that the axial $U(1)_A$ symmetry is embedded in the
chiral $U(N_f)_L\times U(N_f)_R$ symmetry
as a subgroup whose element is of the form
$(e^{i\alpha},e^{-i\alpha})$ which is interpreted
as the gauge transformation with
$\Lambda|_{z=\pm\infty}=\pm i\,\alpha\cdot 1_{N_f}$
in the D4/D8 model. 
Therefore, the theta parameter
is shifted as
\begin{eqnarray}
\delta_\Lambda\theta=2N_f \alpha
\end{eqnarray}
under the $U(1)_A$ symmetry transformation, as expected.

Because $F_2=dC_1$ is not invariant under the gauge
transformation, it is convenient to define
the gauge invariant combination
\begin{eqnarray}
\wt F_2\equiv dC_1 +i\,\tr(A)\wedge\delta(y) dy\ .
\end{eqnarray}
Then the integral of $\wt F_2$ over the $(y,z)$ plane is
\begin{eqnarray}
\int_{\bR^2}\wt F_2=\theta
+i\int_{-\infty}^\infty dz\tr(A_z)
=\theta+\frac{\sqrt{2N_f}}{f_\pi}\eta'\ .
\label{intwtF}
\end{eqnarray}
Here, the $\eta'$ meson is the $U(1)$ part of the pion matrix $\pi$, 
and we have used (\ref{Upi}) and (\ref{pion}) to obtain the relation\
\footnote{We use the $U(N_f)$ generators $T^a$ normalized as
$\Tr T^a T^b=\half\delta_{ab}$, whose $U(1)$ part is 
$T^0=\frac{1}{\sqrt{2N_f}}1_{N_f}$.}
\begin{eqnarray}
i\int_{-\infty}^{\infty}dz\, \tr(A_z)=\frac{2}{f_\pi}
 \tr(\pi)
=\frac{\sqrt{2N_f}}{f_\pi}\eta'\ .
\end{eqnarray}

In the supergravity action, the kinetic term of $C_1$
should be modified in a gauge invariant way as
\begin{eqnarray}
S^{\rm kin}_{C_1}=-\frac{1}{4\pi(2\pi l_s)^6}
\int d^{10}x\sqrt{-g}\,|\wt F_2|^2 \ .
\label{kinC1}
\end{eqnarray}
The solution of the equation of motion
satisfying the condition (\ref{intwtF}) can be found in
Refs.~\citen{Witten;theta} and \citen{Barbon}.
It is given by
\begin{eqnarray}
\wt F_2=\frac{c}{U^4}
\left(\theta+\frac{\sqrt{2N_f}}{f_\pi}\eta'\right)
dU\wedge d\tau \ ,
\label{wtFsol}
\end{eqnarray}
where
\begin{eqnarray}
c\equiv\frac{3\Ukk^3}{\delta\tau}
=\frac{4}{3^5\pi}(\gym^2 N_c)^3 \Mkk^4 l_s^6\ .
\end{eqnarray}
Inserting (\ref{wtFsol}) into the action (\ref{kinC1}),
we obtain
\begin{eqnarray}
S_{C_1}^{\rm kin}=-\frac{\chi_g}{2}\int d^4 x
\left(\theta+\frac{\sqrt{2N_f}}{f_\pi}\eta'\right)^2 \ ,
\label{etamassterm}
\end{eqnarray}
where
\begin{eqnarray}
\chi_g=
\frac{1}{4(3\pi)^{6}}(\gym^2N_c)^3 \Mkk^4
\label{chig}
\end{eqnarray}
is the topological susceptibility.\cite{Witten;theta,Hashimoto-Oz}
This is the mass term of the $\eta'$ meson
with the mass given by the Witten-Veneziano formula \cite{W;u1,V;u1}
\begin{eqnarray}
m_{\eta'}^2=\frac{2N_f}{f_\pi^2}\chi_g\ .
\end{eqnarray}
Inserting (\ref{fpi}) and (\ref{chig}) into this formula,
we obtain
\begin{eqnarray}
m_{\eta'}=\frac{1}{3\sqrt{3}\pi}\sqrt{\frac{N_f}{N_c}}\,
(\gym^2N_c) \Mkk\ .
\end{eqnarray}

\section{Conclusion and discussion}
\label{concdisc}

In this paper, we have discussed aspects of massless QCD
using the D4/D8 model in the probe approximation in which $N_f\ll N_c$.
We have argued that this model reproduces various low-energy
phenomena of massless QCD. 
In particular, the spontaneous breaking of the chiral
$U(N_f)_L\times U(N_f)_R$ symmetry is realized through a smooth
interpolation of D8-\AD 8 pairs in the D4/D8 model.
By analyzing fluctuations around the probe D8-branes,
we have found massless pseudo-scalar mesons which should be identified
with the NG bosons associated with the chiral symmetry breaking.
The low-energy effective action for the pion field on the D8-brane
is consistently written in the form of the chiral Lagrangian.
In fact, approximating the non-Abelian DBI action
of the D8-brane by the Yang-Mills action,
we have shown that the low-energy effective action for the pion
field is actually the same as that in the Skyrme model.

In addition to the massless pion, we have also discussed vector mesons.
An intriguing point here is that the pion and all the
vector mesons result from the same gauge field on the probe D8-brane.
The effective action of these mesons can be elegantly expressed as
that of the five-dimensional Yang-Mills theory in a curved background
(\ref{multiFF}) with the CS-term (\ref{D8CS}).
Inserting the mode expansion of the
gauge potential, we obtain a rather conventional
four-dimensional effective action in terms of 
the pion and the vector mesons.
It turns out that this effective action is closely related to that
obtained in the hidden local symmetry approach.
We evaluated some couplings among the pion and
the lightest vector meson and found that they satisfy a
KSRF-type relation, suggesting that the holographic description
is effective.

We also found that the CS term on the probe D8-brane yields
the WZW term. The WZW term for the pion effective action
is written by using a five-dimensional
manifold whose boundary is our four-dimensional world. This
can be viewed as a prototype of the more recent idea of the
holographic description of four-dimensional gauge theories.
In the D4/D8 model, the five-dimensional manifold
has the physical meaning of the world-volume of the probe D8-brane.
Furthermore, our derivation of the WZW term including the background
gauge potential is even practically simpler than the original one.

We have also studied baryons composed of massless quarks.
As argued by Witten \cite{Witten;baryon},
a baryon can be constructed in the SUGRA
description by introducing a baryon vertex. In the present context,
this is given by a D4-brane wrapped around the $S^4$. In the absence
of the flavor branes, it was shown that $N_c$ fundamental strings
have to be attached to the D4-brane, suggesting that
$N_c$ quarks are bound to form a baryon.
In the D4/D8 model, this wrapped D4-brane is
interpreted as the Skyrmion constructed on the probe D8-brane.
This interpretation helps us understand the relation between
a bound state of $N_c$ quarks and a soliton of
the pion effective theory.

Finally, we investigated how the $U(1)_A$ anomaly can be seen
in the D4/D8 model. We showed that under the $U(1)_A$ transformation,
the $\theta$ angle is shifted in a manner consistent with a field
theory result.
Furthermore, we derived the Witten-Veneziano formula for the mass of
the $\eta'$ meson. Crucial in this derivation is the fact that
the RR 1-form field in the presence of the D8-branes is not gauge
invariant, and the gauge field on the D8-brane
has to couple with the RR 1-form field
in such a way that the RR 1-form field has a gauge invariant
kinetic term. The advantage of our approach is that the $\eta'$ meson
exists in the gauge field on the D8-brane, and we can use the
standard anomaly analysis to derive the coupling without performing an
explicit calculation of the string theory amplitudes. Note also that
the particles created from the fluctuations of the
RR 1-form field are interpreted as glueballs.
Hence, the SUGRA description enables us to calculate
the coupling between the glueballs and the $\eta'$ meson
explicitly, as is partly done in Ref.~\citen{Barbon} for the D4/D6 model.

With all this success, we have come to believe that the D4/D8 model is in the
same universality class as QCD. However,
as mentioned in the introduction, our D4/D8 model
deviates from QCD at the energy scale of $\Mkk$, which is, unfortunately,
the same energy scale of the mass of the vector mesons.
In addition, the supergravity approximation and the probe
approximation may not be sufficiently precise to be applied to
the realistic situation in which $N_c=3$ and $N_f=2$. Nevertheless,
the numerical results for the spectrum of vector mesons
agree well with the experimental data.
Although it is difficult to justify this agreement,
it encourages us to believe that the D4/D8 model successfully
captures even quantitative features of QCD.

We end this paper with some comments on future directions.
{}First, it would be interesting to introduce massive flavors into
the D4/D8 model.
One outcome of obtaining a flavor with
non-vanishing mass is that with it one can compute the chiral
condensate by differentiating the effective action with respect
to the mass parameter, as was done in the D4/D6 model \cite{kmmw}.
One possible way to introduce the quark mass term is to include
the tachyon field of the D8-\AD8 system, as discussed in
Ref.~\citen{SugTak}
for the case of QED.
It is, however, not clear if the effective theory including
the tachyon field is tractable.

It is important to investigate the WZW term considered in \S
\ref{wzwterm} in more detail.
A generalization of the WZW term to incorporate the $\rho$ meson
is carried out in Refs.~\citen{KRS;WZW} and \citen{Fujiwara:1984mp}.
The expression (\ref{CSWZW}) also
defines a systematic way to include couplings to the
vector mesons into the WZW term.
This approach may be even more powerful
than the others, as (\ref{CSWZW}) contains
not only the lightest vector meson (the $\rho$ meson)
but also heavier vector and axial-vector mesons.

The couplings among the mesons and the background gauge fields
are of great interest.
It would be worthwhile to include the background gauge field
in the effective action, as we did for the WZW term, and check
if the vector meson dominance hypothesis holds in
the D4/D8 model. (See Ref.~\citen{HoYoSt} for recent developments along
this line in the context of AdS/CFT.)

We have argued that the five-dimensional Yang-Mills theory
(\ref{multiFF}), which comes from the nine-dimensional non-Abelian
DBI action on the probe D8-brane, gives a simple and powerful
holographic description of the low-energy effective theory of QCD.
In fact, it automatically incorporates the contributions from an
infinite number of scalar and vector mesons.
As a further application, it would be interesting to
study the properties of the instantons considered in \S \ref{baryon}
in more detail to gain deeper insight into baryons.

We can also analyze massless QCD at finite temperatures
using the D4/D8 model. 
{}Following the idea presented in Refs.~\citen{Witten:1998qj} and 
\citen{Witten;thermal},
the authors of Ref.~\citen{kmmw} analyzed the finite temperature
QCD using two SUGRA backgrounds, which are
believed to describe QCD (without flavors) in
high and low temperature phases.
The low temperature solution is obtained from
the solution (\ref{D4sol}) through the Wick rotation $t\ra t_E=it$
and making the Euclidean time periodic as $t_E\sim t_E+ 1/T$.
The high temperature solution is obtained by exchanging
the roles of the $\tau$ and $t_E$ directions in the low temperature
solution.
In Ref.~\citen{kmmw}, the flavor degrees of freedom are obtained
by embedding probe D6-branes into these backgrounds,
and the phase structure of the finite temperature
QCD with flavors is examined.
It is straightforward to extend this study to the D4/D8 model 
by placing a D8-brane probe in the same SUGRA backgrounds.
In this system, we can easily understand the restoration of chiral
symmetry at high temperatures.
Recall that the spontaneous breaking of chiral
 $U(N_f)_L\times U(N_f)_R$ symmetry
is realized as a smooth interpolation of D8-\AD 8 pairs. 
However, this never happens at high temperatures.
This is because in the SUGRA background that governs the high temperature
phase, the circle transverse to the D8-\AD 8 pairs does not
shrink to zero, and therefore the D8-\AD 8
pairs do not intersect each other.
A thorough investigation of finite temperature QCD using the
D4/D8 model would be worthwhile.

\section*{Acknowledgements}

We would like to thank our colleagues at the Yukawa Institute for
Theoretical Physics and the particle theory group
at Ibaraki University for discussions and encouragement. 
S.S. is especially grateful to 
H. Fukaya, Y. Imamura, Y. Kikukawa,
T. Kugo, H. Kunitomo, T. Onogi, K. Takahashi,
M. Tanabashi and K. Yamawaki
for various useful and enjoyable discussions.
T.S. would like to thank the members of the Yukawa Institute
for Theoretical Physics, where this work was initiated, 
for kind hospitality and a stimulating atmosphere.
This work is supported by the Grant-in-Aid for the 21st Century COE
``Center for Diversity and Universality in Physics'' from the Ministry
of Education, Culture, Sports, Science and Technology (MEXT) of Japan.
The work of S.S. is partly supported by
a Rikkyo University Special Fund for Research.

\appendix

\section{Normalization of the RR fields and the Chern-Simons term}
\label{AppCS}

Here we fix our normalization of the RR fields.
A standard CS coupling on a D$p$-brane
is (see Ref.~\citen{Polchinski:1998rq})
\begin{equation}
S_{\rm CS}^{Dp}=\mu_p \int_{Dp} \sum_q C_{q+1}\wedge
 \tr e^{2\pi\alpha^{\prime}F} 
=\mu_p\int_{Dp}\sum_{n=0,1,\cdots} C_{p-2n+1}\wedge
\frac{1}{n!}(2\pi\alpha^{\prime})^n\tr F^n\ ,
\label{CS:standard}
\end{equation}
with $\mu_p=(2\pi)^{-p}l_s^{-(p+1)}$.
Recall that $\mu_p$ is fixed from the following action of $C_{p+1}$
with a minimal coupling constant given by $\mu_p$:
\begin{equation}
S_{\rm RR}=-\frac{1}{4\kappa_{10}^2}\int_{10}
 F_{p+2}\wedge^{\ast}F_{p+2}
+{\mu}_p\int_{Dp} C_{p+1} \ ,
\end{equation}
where $2\kappa_{10}^2=(2\pi)^7 l_s^8$ and $F_{p+2}=dC_{p+1}$.
The equation of motion for $C_{p+1}$ is given by
\begin{equation}
\frac{1}{2\kappa_{10}^2}d^{\ast}F_{p+2}=\mu_p\delta_{\bot} \ .
\end{equation}
Thus a unit electric charge of $C_{p+1}$ takes the form
\begin{equation}
\int_{S^{8-p}}{}^{\ast}F_{p+2}=\int_{S^{8-p}}dC_{7-p}
=2\kappa_{10}^2\mu_p  \ .
\end{equation}
By rescaling $C_{p+1}$ as
\begin{equation}
C_{p+1} \rightarrow \frac{\kappa_{10}^2\mu_{6-p}}{\pi} C_{p+1} \ ,
\label{rescale}
\end{equation}
an RR charge is measured in units of $2\pi$.
This is the convention employed in this paper.
In this case, the minimal coupling of an RR potential becomes
\begin{equation}
\mu_p\frac{\kappa_{10}^2\mu_{6-p}}{\pi}\int_{Dp} C_{p+1}
=\int_{Dp} C_{p+1} \ .
\label{minimal}
\end{equation}
Using (\ref{rescale}), (\ref{CS:standard}) can be rewritten as
\begin{eqnarray}
S^{Dp}_{\rm CS}
&\!=\!&
\int_{Dp}\sum_{n=0,1,\cdots} 
\frac{\kappa_{10}^2\mu_p\mu_{6-p+2n}}{\pi}\,\,(2\pi\alpha^{\prime})^n\,
C_{p-2n+1}\wedge\frac{1}{n!}\tr F^n \nn \\
&\!=\!&
\int_{Dp}\sum_{n=0,1,\cdots} 
C_{p-2n+1}\wedge\frac{1}{(2\pi)^n n!}\tr F^n \nn \\
&\!=\!&
\int_{Dp} 
\sum_qC_{q+1}\wedge\tr e^{\frac{F}{2\pi}} \ .
\end{eqnarray}
In this convention, $C_{p+1}$ and $C_{7-p}$ are related
by the Hodge dual as
\begin{equation}
{}^{\ast}dC_{p+1}=\frac{\mu_p}{\mu_{6-p}}\, dC_{7-p}
=(2\pi l_s)^{2(3-p)}\,dC_{7-p} \ .
\end{equation}
Also, the kinetic term of $C_{p+1}$ becomes
\begin{equation}
S_{\rm RR}^{\rm kin}=-\frac{1}{4\pi} (2\pi l_s)^{2(p-3)}
\int_{10} dC_{p+1}\wedge^{\,\ast}\!dC_{p+1} \ .
\end{equation}

\section{Fluctuations of fermions on D8}
\label{fermion}
Here we consider a single fermion on a probe D8-brane.

We start with the action
\footnote{Fermions on D-branes in a generic SUGRA background
couple to the dilaton and RR-potentials \cite{Marolf:2003ye,Marolf:2003vf}.
In the present case, it is not difficult to show that their
contribution to the action (\ref{action;f}) vanishes.
}
\begin{equation}
S_{D8}^f=k\int d^9x\,\sqrt{-\det g}\,\,e^{-\phi}\,\bar{\Psi}\,
i\Dsl_{\,9}\Psi \ ,
\label{action;f}
\end{equation}
where $k$ is a constant,
$\Psi$ is a Majorana spinor, and
$g$ is the (8+1)-dimensional induced metric on the
probe D8-brane given by
\begin{equation}
ds_9^2=(R^3\Ukk)^{1/2}\left[
\frac{4}{9}\Mkk^2\,K^{1/2}dx_{\mu}^2
+\frac{4}{9}K^{-5/6}dZ^2+K^{1/6}d\Omega_4^2
\right] \ .
\label{indmetric}
\end{equation}
The Dirac operator $i\Dhtsl_9$ of the metric (\ref{indmetric})
can be evaluated as
\begin{equation}
i\Dhtsl_9=\frac{1}{(R^3\Ukk)^{1/4}}\left[
K^{-1/4}\,i\delsl
+\frac{2}{3}\Mkk K^{-1/12}\,i\Dhtsl
+\Mkk\Gamma_4\left(K^{5/12}\,i\del_Z+\frac{4}{3}\,iZK^{-7/12}
\right)\right] \ .
\end{equation}
Here we define $A=(a,4,l)$ as an index of
a (8+1)-dimensional local Lorentz frame,
with $a=0,1,2,3$ and $l=5,6,7,8$.
$\Sigma_{AB}=\frac{1}{4}[\Gamma_A,\Gamma_B]$ is the generator of
$SO(8,1)$ in the spin representation.
$\delsl=\Gamma^{\mu}\del_{\mu}$ and
$\Dhtsl=\Gamma^l\Dht_l$ 
with
$\Dht_l=\hat{\del}_l+\wht^{mn}_{~~~,l} \Sigma_{mn}$,
which is written in the ortho-normal frame of a unit $S^4$.

Note that an (8+1)-dimensional gamma matrix satisfies the relation
\begin{equation}
\cC^T=\cC \ , ~~~ (\Gamma^{A})^T=\cC^{-1}\Gamma^A\cC \ ,
\end{equation}
where $\cC$ is a charge conjugation matrix.
{}From this and the fact that $\Psi$ is a Majorana spinor, we find
\begin{equation}
\bar{\Psi}\Gamma_A\Psi=0 \ .
\label{zero}
\end{equation}
Then, rescaling $\Psi$ by $\wt\Psi$ given by
\begin{equation}
\Psi=K^{-13/24}\,\wt\Psi \ ,
\end{equation}
and using (\ref{zero}), the action becomes
\begin{equation}
S_{D8}^f=\kt \int d^4x\,dZ\,\epsilon_4\,\bar{\wt\Psi}
\left[
K^{-2/3}i\delsl + \Mkk\left(
i\Gamma_4\del_Z+\frac{2}{3}K^{-1/2}i\Dhtsl\right)\right]
\wt\Psi \ ,
\label{action;Psi}
\end{equation}
with $\kt=\frac{2k}{3}\frac{\Ukk}{g_s}\UoR^{15/4}$.
The Dirac equation for $\wt\Psi$ reads
\begin{equation}
\left[
K^{-2/3}i\delsl+\Mkk\left(
i\Gamma_4\del_Z+\frac{2}{3}K^{-1/2}i\Dhtsl\right)
\right]\wt\Psi=0 \ .
\label{eom;Psi}
\end{equation}

{}From the fact that $SO(8,1)$ possesses $SO(3,1)\times SO(4)$ as a subgroup,
the $SO(8,1)$ gamma matrix can be realized as
\begin{equation}
\Gamma^a=\gamma^a \otimes I_4 \ , ~~~
\Gamma^4=\bar{\gamma} \otimes \bar{\rho} \ ,~~~
\Gamma^l=\bar{\gamma} \otimes \rho^l \ .
\end{equation}
Here, $\gamma^a$ is the $SO(3,1)$ gamma matrix, and
$\rho^l$ is the $SO(4)$ gamma matrix. 
$\bar{\gamma}$ and $\bar{\rho}$ are the chirality matrices given by
$\bar{\gamma}=i^{-1}\gamma^0\gamma^1\gamma^2\gamma^3$,
$\bar{\rho}=\rho^5\rho^6\rho^7\rho^8$.
Note that $\cC$ can be written as
\begin{equation}
\cC=C\otimes C^{\prime} \ ,
\end{equation}
where $C$ and $C^{\prime}$ are the $SO(3,1)$ and $SO(4)$ charge conjugation
matrices.
Multiplying (\ref{eom;Psi}) by $\bar{\gamma}\otimes I_4$, we obtain
\begin{equation}
\left[
K^{-2/3}\,i\bar{\gamma}\delsl_4\otimes I_4
+\Mkk\,I_4\otimes
\left(i\bar{\rho}\del_Z+\frac{2}{3}K^{-1/2}i\Dhtsl_4\right)
\right]\wt\Psi=0 \ .
\end{equation}
Here, we have $\delsl_4=\gamma^{\mu}\del_{\mu}$, and $\Dhtsl_4$ is the Dirac
operator of a unit $S^4$.

Let us assume that $\wt\Psi$ takes the form
\begin{equation}
\wt\Psi(x^{\mu},Z,S^4)=
\sum_n\left( \psi_n(x^{\mu})\wt\Psi_n(Z,S^4)
+\psi_n^c(x^{\mu})\wt\Psi_n^c(Z,S^4)\right) \ ,
\label{ansatz;Psi}
\end{equation}
where $\psi_n(x^{\mu})$ is a Dirac spinor of $SO(3,1)$ and
$\psi_n^c=C\bar{\psi}_n^T$.
Also, we have
\begin{eqnarray}
\wt\Psi_n(Z,S^4)&\!=\!&
f_+^{(n)}(Z)\chi_+(S^4)+f_-^{(n)}(Z)\chi_-(S^4) \ , \nn\\
\wt\Psi_n^c(Z,S^4)
&\!=\!&
f_+^{(n)\ast}(Z)\chi_+^c(S^4)+f_-^{(n)\ast}(Z)\chi_-^c(S^4)
\ . 
\end{eqnarray}
Here, the function $\chi_{\pm}$ are given by
$\chi_{\pm}=\frac{1}{2}(1\pm\bar{\rho})\chi$, where $\chi$ is
the eigenfunction of $\Dhtsl_4$ satisfying
\begin{equation}
i\Dhtsl_4\,\chi=\beta\chi \ ,
\end{equation}
and \begin{equation}
\chi^c=C^{\prime}\chi^{\ast} \ ,~~~
\chi_{\pm}^c=\frac{1}{2}(1\pm\bar{\rho})\chi^c \ .
\end{equation}
Note that $\chi^c$ satisfies $i\Dhtsl_4\,\chi^c=-\beta\chi^c$.
To fix $f^{(n)}_{\pm}$, we require 
\begin{equation}
\left(
i\bar{\rho}\del_Z+\frac{2}{3}K^{-1/2}i\Dhtsl_4\right)\wt\Psi_n
=\lambda_n^{\prime\prime}K^{-2/3}\wt\Psi_n \ .
\end{equation}
We can then verify that the functions $f_{\pm}^{(n)}$ satisfy
\begin{eqnarray}
i\del_Zf_+^{(n)}+\frac{2}{3}\beta K^{-1/2}f_-^{(n)}
-\lambda_n^{\prime\prime}K^{-2/3}f_+^{(n)}
&\!=\!&0 \ , 
\nn\\
-i\del_Zf_-^{(n)}+\frac{2}{3}\beta K^{-1/2}f_+^{(n)}
-\lambda_n^{\prime\prime}K^{-2/3}f_-^{(n)}
&\!=\!&0 \ .
\label{f+f-}
\end{eqnarray}
We normalize $f_{\pm}^{(n)}$ as
\begin{equation}
\int dZ K^{-2/3}\,f_+^{(m)\ast}f_+^{(n)}=
\int dZ K^{-2/3}\,f_-^{(m)\ast}f_-^{(n)}=\frac{1}{4\wt k}\,\delta_{mn} \ .
\label{norm;f}
\end{equation}
It can then be verified that $\wt\Psi_n$ satisfy the following
normalization conditions:
\begin{eqnarray}
&&\int dZK^{-2/3}\,\epsilon_4
\left(\wt\Psi_n^c\right)^TC^{\prime -1}\wt\Psi_m=
-\int dZK^{-2/3}\,\epsilon_4\,
\wt\Psi_m^T\,C^{\prime -1}\wt\Psi_n^c=\frac{1}{2\wt k}\,\delta_{mn}
\ , \nn\\
&&\int dZK^{-2/3}\,\epsilon_4\wt\Psi_n^T\, C^{\prime -1}\wt\Psi_m=
\int dZK^{-2/3}\,\epsilon_4\left(\wt\Psi_n^c\right)^TC^{\prime -1}
\wt\Psi_m^c=0 \ .
\label{norm;Psi}
\end{eqnarray}
The normalization condition (\ref{norm;f}) is satisfied if
$f_\pm^{(n)}$ behaves as $f_\pm^{(n)}\sim\cO(Z^{a})$
with $a<1/6$ for $Z\ra\infty$.
It is easy to determine the asymptotic behavior of the solution of
(\ref{f+f-}). We find
\begin{equation}
f_+^{(n)}\sim Z^{\pm\frac{2\beta}{3}} \ , ~~~
f_-^{(n)}\sim \mp iZ^{\pm\frac{2\beta}{3}} \ .~~~~
(\mbox{as}~|Z|\ra\infty)
\end{equation}
Solutions with negative exponents represent normalizable modes.

Using (\ref{norm;Psi}), we end up with an action of the form
\begin{equation}
S_{\rm D8}^f=\sum_n\int d^4x\Big(
-i\bar{\psi}_n\delsl_4\psi_n
-\lambda_n''\Mkk\,\bar{\psi}_n\bar{\gamma}\psi_n
\Big) \ .
\end{equation}
Thus, $\psi_n$ represents a fermionic meson that has no
counterpart in QCD.


%

\end{document}